\documentclass[sigconf]{acmart}

\AtBeginDocument{%
  }

\copyrightyear{2024}
\acmYear{2024}
\setcopyright{rightsretained}
\acmConference[RecSys '24]{18th ACM Conference on Recommender Systems}{October 14--18, 2024}{Bari, Italy}
\acmBooktitle{18th ACM Conference on Recommender Systems (RecSys '24), October 14--18, 2024, Bari, Italy}\acmDOI{10.1145/3640457.3688121}
\acmISBN{979-8-4007-0505-2/24/10}

\usepackage{enumitem}
\usepackage{makecell}
\usepackage{color, colortbl, xcolor}
\usepackage{multirow}
\usepackage{listings}
\usepackage{fancybox}
\usepackage{float}
\definecolor{Gray}{gray}{0.92}

\lstdefinelanguage{json}{
    basicstyle=\scriptsize\ttfamily,
    showstringspaces=false,
    breaklines=true,
    frame=trbl,
}

\usepackage{algorithmic}
\usepackage[ruled, linesnumbered, vlined, boxed]{algorithm2e} 

\newcommand{\rparagraph}[1]{\vspace{2.0mm}\noindent\textbf{#1.}}
\newcommand{\rparagraphnodot}[1]{\vspace{2.0mm}\noindent\textbf{#1}}
\newcommand{\sparagraph}[1]{\vspace{2.0mm}\noindent\textbf{#1.}}

\newcommand{\RN}[1]{%
  \textup{\uppercase\expandafter{\romannumeral#1}}%
}
\definecolor{mygray}{rgb}{0.8,0.8,0.8}

\begin{document}

\title{CALRec: \underline{C}ontrastive \underline{A}lignment of Generative \underline{L}LMs for Sequential \underline{Rec}ommendation}


\author{Yaoyiran Li}
\authornote{Both authors contributed equally to this research. Corresponding authors: Yaoyiran Li \{yl711@cam.ac.uk\} and Xiang Zhai \{xiangzhai@google.com\}.}
\authornote{This work has been done during the author’s internship at Google.}
\affiliation{%
   \institution{University of Cambridge}
   \country{United Kingdom}
}
\email{yl711@cam.ac.uk}

\author{Xiang Zhai}
\authornotemark[1]
\affiliation{%
   \institution{Google}
   \country{USA}
}
\email{xiangzhai@google.com}

\author{Moustafa Alzantot}
\affiliation{%
   \institution{Google}
   \country{USA}
}
\email{mfarid@google.com}

\author{Keyi Yu}
\affiliation{%
   \institution{Google}
   \country{USA}
}
\email{yukeyi@google.com}

\author{Ivan Vulić}
\affiliation{%
   \institution{University of Cambridge}
   \country{United Kingdom}
}
\email{iv250@cam.ac.uk}

\author{Anna Korhonen}
\affiliation{%
   \institution{University of Cambridge}
   \country{United Kingdom}
}
\email{alk23@cam.ac.uk}

\author{Mohamed Hammad}
\affiliation{%
   \institution{Google}
   \country{USA}
}
\email{mibra@google.com}











\renewcommand{\shortauthors}{Li et al.}

\begin{abstract}
Traditional recommender systems such as matrix factorization methods have primarily focused on learning a shared dense embedding space to represent both items and user preferences. Subsequently, sequence models such as RNN, GRUs, and, recently, Transformers have emerged and excelled in the task of sequential recommendation. This task requires understanding the sequential structure present in users’ historical interactions to predict the next item they may like. Building upon the success of Large Language Models (LLMs) in a variety of tasks, researchers have recently explored using LLMs that are pretrained on vast corpora of text for sequential recommendation. To use LLMs for sequential recommendation, both the history of user interactions and the model’s  prediction of the next item are expressed in text form. We propose CALRec, a two-stage LLM finetuning framework that finetunes a pretrained LLM in a two-tower fashion using  a mixture of two contrastive losses and a language modeling loss: the LLM is first finetuned on a data mixture from multiple domains followed by another round of target domain finetuning. Our model significantly outperforms many state-of-the-art baselines ($+37\%$ in Recall@1 and $+24\%$ in NDCG@10) and our systematic ablation studies reveal that (i) both stages of finetuning are crucial, and, when combined, we achieve improved performance, and (ii) contrastive alignment is effective among the target domains explored in our experiments.
\end{abstract}

\ccsdesc[500]{Information systems~Recommender systems}
\ccsdesc[500]{Computing methodologies~Neural networks}

\keywords{Sequential Recommendation,
Large Language Models,
Contrastive Learning}


\maketitle

\section{Introduction and Motivation}
\label{s:introduction}
Recommender systems aim to understand users and items and use the learned knowledge to recommend relevant items for future user interactions. Traditional collaborative filtering methods~\cite{10.1145/138859.138867,10.1145/371920.372071,5197422} map item and user IDs into a common latent embedding space learned from individual user-item interactions, but these embeddings are often static and fail to capture the dynamic nature of user interests. Recent exploration in sequential recommender employs RNN~\cite{hidasi2016sessionbased}, CNN~\cite{yan2019cosrec} and the transformer~\cite{8594844, 10.1145/3357384.3357895} to directly learn the time-evolution of user behaviors. The transformer architecture ~\cite{NIPS2017_3f5ee243, devlin-etal-2019-bert} stands out for its efficient capability of modeling sequences through the self-attention mechanism, which learns the correlation or causal relationships between different tokens and their influence on future tokens. In a large body of recommender systems (RecSys) research, each input position (token) corresponds to a single user activity, typically a user's interaction (e.g., purchase) with an item ~\cite{8594844, 10.1145/3357384.3357895, ijcai2019p600, 10.1145/3340531.3411954, 10.1145/3534678.3539381}.
More specifically, some approaches rely on item IDs, each represented by a unique token, and model a user's interaction history as a sequence of item ID tokens~\cite{8594844,10.1145/3357384.3357895}; other efforts combine more sophisticated features, such as the text embedding of item descriptions, to capture richer content information for each item ~\cite{ijcai2019p600,10.1145/3340531.3411954}.

Motivated by the global paradigm shift towards autoregressive Language Large Models (LLMs)~\cite{2023arXiv230308774O, anil2023palm,tay2023ul, 2023arXiv231211805G} with successful applications in various tasks and domains~\cite{DBLP:conf/iclr/DosovitskiyB0WZ21, 9710415, 2022arXiv220406125R, yu2021vector}, we focus on leveraging generative LLMs for sequential recommender systems. Pretrained LLMs learn knowledge from large text corpora and are expected to perform reasonably well in general tasks, and their performance on specific tasks further improves after fine-tuning with domain-specific data. Several studies have explored utilizing 
LLMs in recommender systems~\cite{2023arXiv230519860W}, such as framing sequential recommendation as sentence-completion~\cite{li2023gpt4rec} or question-answering problems~\cite{zhang2023recommendation}. 



Despite the impressive language understanding capabilities of pretrained LLMs, they require fine-tuning for highly specific tasks like sequential recommendation. An LLM should be ideally fully fine-tuned~\cite{2023arXiv230506474K, 2023arXiv231108572W} to learn to comprehend domain-specific flattened text sequences and the underlying user interest evolution. Similar to most sequence modeling approaches like SASRec~\cite{8594844}, BERT4REC~\cite{10.1145/3357384.3357895}, Recformer~\cite{10.1145/3580305.3599519}, and GPT4Rec~\cite{li2023gpt4rec}, we adopt next-item prediction as the sequential recommendation task. 

Based on sentence-completion~\cite{li2023gpt4rec} or question-answering ~\cite{zhang2023recommendation} styled prompting, our template design is further inspired by the idea of few-shot learning~\cite{brown2020language}. Specifically, we prefix user sequence and each item in the user sequence with distinctive text indicators that \textbf{1)} allow the model to better understand data pattern and text format and \textbf{2)} increase the LLM response coherence of the input sentence. We further propose a Contrastive Aligned Generative LLM Recommendation (CALRec) framework to adapt an LLM for sequential recommendation, inspired by the effectiveness of contrastive learning (CL) in mapping-based RecSys~\cite{10.1145/3534678.3539381,10.1145/3580305.3599519} and CL research in NLP and other areas ~\cite{gao-etal-2021-simcse,li-etal-2023-translation}. CALRec features the following:

\begin{itemize}
\item Pure text input and text output with advanced prompt design. 
\item A mixed training objective that combines customized next-item generation objective and auxiliary contrastive objectives.
\item A two-stage LLM fine-tuning paradigm consisting of a multi-category joint fine-tuning stage followed by a category-specific fine-tuning stage.
\item A novel quasi-round-robin BM25 retrieval approach for item retrieval.
\end{itemize}


Similar to previous work~\cite{8594844,10.1145/3357384.3357895,10.1145/3580305.3599519}, we leverage the Amazon Review dataset~\cite{ni-etal-2019-justifying} and compare our results to open-source state-of-the-art (SotA) baselines. We observed a non-negligible percentage of consecutive identical activities by the same users in the raw datasets which might trivialize the next-item prediction task. We thus train and evaluate the model on deduplicated sequences and propose to adopt `Last Item Repeater', a training-free, text-statistic baseline for model comparison (\S\ref{s:expsetup} and \S\ref{s:discussion}). 
We also revisit the existing evaluation metrics of the previous works, and propose to report both optimistic and pessimistic metric scores for text-based RecSys (\S\ref{s:expsetup}). Through comprehensive experiments, we discover that our CALRec method demonstrates superior performance and outperforms existing SotA by a considerable margin in most evaluation metrics. A systematic ablation study indicates the effectiveness of the two-stage training paradigm, our contrastive alignment, and our template design among the various data categories examined in our experiments. 

Finally, we briefly summarize the contribution of our study:
\begin{itemize}[leftmargin=*]
\item We proposed CALRec, a novel sequential recommendation framework that features 
advanced prompt design, 
a two-stage training paradigm, a combined training objective, and a quasi-round-robin BM25 retrieval approach.
\item We conduct comprehensive experiments, ablation study and further analyses and show that \textbf{1)} our approach outperforms existing baselines 
in most metrics by a considerable margin; \textbf{2)} the main components of our approach are effective. We also discuss \textbf{3)} the characteristics of our output and our limitations.    
\item We revisit data preprocessing and evaluation metrics in this field.
\end{itemize}

\section{Related Work}
\label{s:relatedwork}
Traditional recommender systems like Collaborative Filtering~\cite{10.1145/138859.138867} learn relevance through user-item co-occurrence. Early collaborative filtering approaches like Matrix Factorization (MF)~\cite{5197422} are ID-only where both user and item IDs are embedded in a shared embedding space. Later MF improvements incorporated other attributes~\cite{10.1145/2556270} such as time and location. Nevertheless, these approaches did not capture the sequential nature of user behavior which gave rise to a class of Sequential Recommenders. The initial Sequential Recommenders, similar to MF, were ID-only models, exemplified by GRU4Rec~\cite{hidasi2016sessionbased}, SASRec~\cite{8594844}, and Bert4Rec~\cite{10.1145/3357384.3357895}. Subsequent research expanded these models by integrating additional attributes beyond IDs, leading to the development of methods such as FDSA~\cite{ijcai2019p600}, S3-Rec~\cite{10.1145/3340531.3411954}, and UniSRec~\cite{10.1145/3534678.3539381}. Due to the challenges associated with IDs in real world Recommender Systems~\cite{10.1145/2556270}, researchers explored removing them in favor of attribute-only models~\cite{10.1145/3534678.3539381,10.1145/3580305.3599519}. However, this shift led to quality issues, as the valuable `memorization characteristics' provided by IDs were lost. Another thread of research explored deriving IDs from attribute embedding~\cite{rajput2023recommender, 2023arXiv230608121S, 10.1145/3543507.3583434}, which mitigates some of the challenges associated with random IDs, albeit at the cost of increased system complexity.


With the advent of pretrained LLMs, which include encoder-only models like BERT~\cite{devlin-etal-2019-bert,10.1145/3580305.3599519}, encoder-decoder models like T5~\cite{10.5555/3455716.3455856}, and decoder-only models like PaLM~\cite{10.5555/3648699.3648939, anil2023palm}, GPT~\cite{brown2020language}, and LLama~\cite{2023arXiv230213971T}, researchers explored leveraging these backbones in sequential recommendation tasks, harnessing the extensive `knowledge' encoded within their parameters. Some approaches utilized zero-shot or few-shot methods~\cite{wang2023zero, 2023arXiv230410149L, hou2024llmrank}, while others extract item text embeddings from LLMs, both with frozen LLM parameters.

Previous work also explored fine-tuning encoder-only, encoder-decoder or decoder-only LLMs to incorporate recommendation knowledge into the LLMs~\cite{10.1145/3580305.3599519,10.5555/3455716.3455856, 2023arXiv230519860W,lin2023can,2023arXiv230506474K, zhang2023recommendation,10.1145/3604915.3608857,10.1007/978-3-031-56063-7_42,10.1145/3589334.3645347}.
P5~\cite{10.1145/3523227.3546767} fine-tuned a pre-trained T5~\cite{10.5555/3455716.3455856} model to solve various common RecSys tasks, including sequential recommendation. In contrast to P5, which frames the sequential recommendation task as an ID generation problem, our approach focuses on generating item attribute text and features a two-tower training framework, enhancing model's understanding of items with their attributes within the context of user-item interaction sequences. GPT4Rec~\cite{li2023gpt4rec} fine-tunes a pre-trained GPT-2 model to generate the next-item titles. Similar to GPT4Rec, we use BM25~\cite{robertson2009probabilistic} to search for items in the item corpus. However, unlike GPT4Rec which was directly fine-tuned for next-item generation as a sentence completion task on each target domain respectively, we are among the first to explore using the contrastive loss as an auxiliary objective to fine-tune an autoregressive LLM on the sequential recommendation task, and our work also features a multi-stage fine-tuning framework, enabling our model to generalize to multiple target domains. We include the list of baseline models we compare against and their details in \S\ref{s:expsetup}.



























\section{Methodology}
\label{s:methodology}
\subsection{Preliminaries and Terminology}

We assume a set of users $\mathcal{U}$, where a user $u\in\mathcal{U}$ is associated with an interaction sequence comprising $n$ items $(I_{1},I_{2},...,I_{n})$ arranged in chronological order (note that the value of $n$ may vary across different users). The \textit{sequential recommendation} task is then defined as follows: given the user’s historical interaction sequence $I_{1:n-1} = (I_{1},I_{2},...,I_{n-1})$, a sequential recommender aims to predict the target item $I_{n}$, the last item in the user's full interaction sequence. In our study, we use pure text to represent each item and user interaction sequence (\S\ref{s:method_template}), and we adapt autoregressive LLMs pretrained on large corpora of text for sequential recommendation. Specifically, we pose the task as a sequence-to-sequence generation task where the input is the text description of $I_{1:n-1}$ and the expected output is that of $I_{n}$. Our primary motivation for formulating the problem entirely within the text domain is to exploit the rich textual information inherent in real-world data and to capitalize on LLMs' strong language understanding and reasoning capabilities.

\begin{figure}[ht!]
    \centering
    \includegraphics[width=1.0\linewidth]{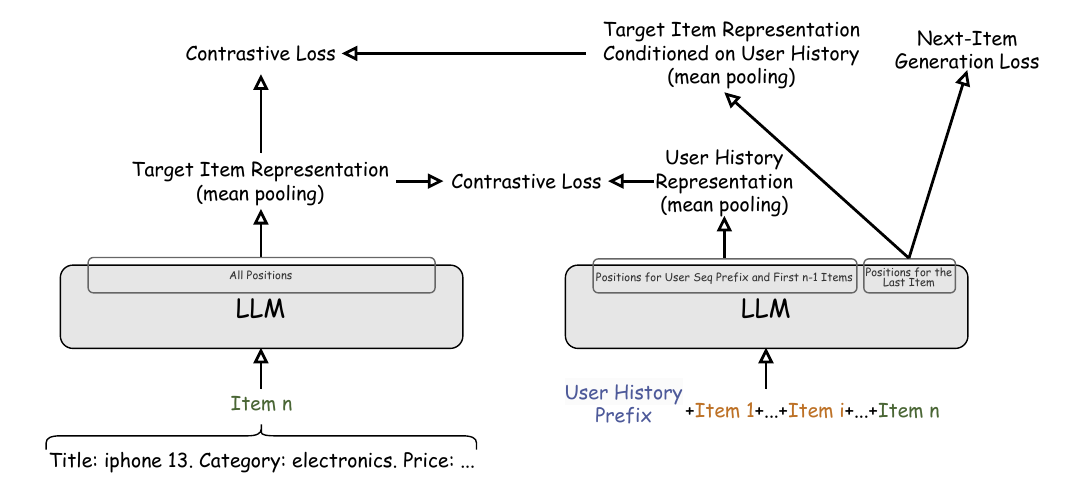}
            \caption{CALRec: Training Framework.}
            \label{fig:framework}
\end{figure}
\subsection{Data Format and Template Design}
\label{s:method_template}
In our setup, items and user sequences are provided exclusively in text format. Instead of treating per-item features separately as in traditional recommender models~\cite{ijcai2019p600,10.1145/3340531.3411954} and inspired by~\citet{10.1145/3580305.3599519}, we flatten each item’s semantically meaningful attributes into a single sequence of text tokens, where the attributes are respectively formatted in the style of ``\texttt{Attribute Name: Attribute Description}'' and then are concatenated together. Typical attributes include the title, ratings, keywords, price, and others, exemplified by the title attribute ``\texttt{Title: iPhone 13 Pro Max}''. We discard non-semantic ID attributes such as item hashing IDs or user IDs which may add noise to the LLM input. Missing attribute descriptions are replaced with the wildcard text ``\texttt{Unknown}''. For the user sequence, different from~\citet{10.1145/3580305.3599519}, we additionally use a user prefix ``\texttt{This is the summary of a user’s purchase history.}'' and per-item prefixes ``\texttt{$\backslash$nThe first/next item bought is as follows.}'' (``\texttt{$\backslash$n}'' is not applied on the first item) to separate items. The user sequence prefix signature facilitates LLMs in differentiating between single item input and user sequence input. The per-item prefix signature
aids LLMs in segmenting distinct items within a user history sequence, thereby enhancing task comprehension. In addition, we end the user history sequence with the same item prefix, prompting the LLM to predict the next purchase as a sentence completion task. This design is inspired by the idea of `few-shot in-context learning'~\cite{brown2020language,li-etal-2023-bilingual,li2024self}:\footnote{Our approach does not conform to the exact concept of `few-shot' learning by definition, since our input data is always from a single user.} since the same item prefix is also seen in the history sequence, our approach is beneficial for the LLM to generate text that follows the output format. A concrete example of the full prompt is provided as follows\footnote{During training, the user history and the target item sequences (from the training set) are concatenated. At inference, only the user history sequence (from the validation/test set) serves as the LLM input.}:
\begin{center}
\fbox{%
\begin{minipage}{0.47\textwidth}
\footnotesize
$\vcenter{\hbox{\rule{0.23\textwidth}{0.5pt}}}$
\textrm{[USER PURCHASE HISTORY SEQUENCE]}
$\vcenter{\hbox{\rule{0.23\textwidth}{0.5pt}}}$
\texttt{\textcolor[RGB]{79, 89, 153}{\newline This is the summary of a user’s purchase history.} The first item bought is as follows. \textcolor[RGB]{189, 113, 38}{Title: The Hillman Group 591520 Small Wire Nail and Brad Assortment, 260-Pack. Category: Tools $\&$ Home Improvement. Brand: Hillman. Price: Unknown.}
\newline The next item bought is as follows. \textcolor[RGB]{189, 113, 38}{Title: Pro Flex Super Flexible Elastomeric Sealant Acrylic Clear Paintable 10 Oz. Category: Tools $\&$ Home Improvement. Brand: Geocel. Price: $\$11.22$.} \newline The next item bought is as follows.\newline}

$\vcenter{\hbox{\rule{0.31\textwidth}{0.5pt}}}$
\textrm{[TARGET ITEM SEQUENCE]}
$\vcenter{\hbox{\rule{0.31\textwidth}{0.5pt}}}$
\texttt{\textcolor[RGB]{74, 102, 53}{\newline Title: Trimaco Llc RF36 35-Inch by 166-Feet Rosin Flooring Paper, Red. Category: Tools $\&$ Home Improvement. Brand: Trimaco. Price: Unknown.}}
\end{minipage}}
\end{center}
Different from our method, many previous text-based approaches do not specify an item prefix and simply leverage a one-off user history suffix~\cite{li2023gpt4rec,zhang2023recommendation} (some approaches even do not specify any user suffix at all~\cite{10.1145/3580305.3599519}). 

\subsection{Two-Stage Fine-Tuning} 
Previous work~\cite{10.1145/3534678.3539381,10.1145/3580305.3599519} indicates that pretraining a RecSys on a large volume of data (various data categories) provides favorable inductive bias for downstream tasks (a specific and smaller-scale data category for evaluation after further fine-tuning on them). Analogously, based on an LLM already pretrained on large, diverse text corpora, we propose a two-stage fine-tuning framework that consists of Multi-Category Joint Fine-Tuning (Stage \RN{1}) and Category-Specific Fine-Tuning (Stage \RN{2}): the latter benefits from the former as a result of transfer learning. Our Stage \RN{1} fine-tuning allows the model to adapt to the sequential recommendation problem setting and learn data patterns in a category-agnostic fashion. Inspired by~\citet{xue-etal-2021-mt5}, we adopt a sampling rate for each data category proportional to $|\mathcal{U}|^{0.3}$, where $|\mathcal{U}|$ is the size (number of users) of each category, for the purpose of combating the data imbalance issue, preventing a few extra-large categories from dominating the Stage \RN{1} fine-tuning set. After deriving the multi-category fine-tuned model, we further adapt the model to each category respectively via Category-Specific Fine-Tuning (Stage \RN{2}). In both stages, the training objective is the same, introduced in \S\ref{s:method_objectives}.

\subsection{Training Objectives}
\label{s:method_objectives}
\rparagraph{Next-Item Generation Objective} The primary objective employed for LLM fine-tuning is the Next-Item (Text) Generation \textit{(NIG)} Objective. NIG aims to generate the text description of the target item given the text description of the previous items in a user's history. Denote the text-tokenized user sequence (for a user) as $\textsc{t}=(t_{1},t_{2},...,t_{l})$, where $l$ is the tokenized sequence length. The first $m$ tokens belong to all the items in the user sequence except for the last item (i.e., the target item), followed by  $l-m$ tokens for the target item. We propose this next-item generation objective to adapt an LLM for sequential recommendation:
\begin{equation}
\label{formula:NIP}
\mathcal{L}_{NIG}=\mathop{\mathbb{E}}\limits_{\textsc{t} \sim p(\textsc{t})}\left[\sum_{j=m+1}^{l}\text{log}\, P(t_{j}|t_{1:j-1};\theta) \right]
\end{equation}
where p(\textsc{t}) denotes the data distribution, and $\theta$ is the whole set of trainable parameters of the chosen LLM.

\rparagraph{Auxiliary Contrastive Alignment} One shortcoming of the NIG is that it deals with the input and output on the token level rather than on the item/user level. Therefore, in addition to the token-level NIG objective, we also investigate auxiliary contrastive objectives that operate on the item/user level. Specifically, we adopt a two-tower training framework where one tower only takes in the target item, and the mean-pooled feature in the final hidden layer is denoted as $\mathbf{v}^{T}$. The second tower takes the whole user sequence and similarly we denote the mean-pooled feature corresponding to the user history sequence as $\mathbf{v}^{U}$ and the feature for the target item conditioned on the user history as $\mathbf{v}^{T|U}$ (see also Fig.~\ref{fig:framework}). We experiment with two contrastive losses for user- and item-level alignments following the InfoNCE loss ~\cite{oord2018representation}, a standard and robust choice in contrastive learning research ~\cite{pmlr-v119-chen20j,gao-etal-2021-simcse,li-etal-2022-improving,li-etal-2023-translation}. Below, we present the calculation of these losses on a batch of data:
%
\begin{align}
\begin{split}\label{formula:CL_TT}
\mathcal{L}_{TT}=&-\frac{1}{N_{b}}\!\sum_{i=1}^{N_{b}}\text{log} \frac{\text{exp}(\text{cos}(\mathbf{v}^{T|U}_{i},\mathbf{v}^{T}_{i})/\tau_{c})}{\sum_{j=1}^{N_{b}}\text{exp}(\text{cos}(\mathbf{v}^{T|U}_{j},\mathbf{v}^{T}_{i})/\tau_{c})},
\end{split}\\
\begin{split}\label{formula:CL_UT}
\mathcal{L}_{UT}=&-\frac{1}{N_{b}}\!\sum_{i=1}^{N_{b}}\text{log} \frac{\text{exp}(\text{cos}(\mathbf{v}^{U}_{i},\mathbf{v}^{T}_{i})/\tau_{c})}{\sum_{j=1}^{N_{b}}\text{exp}(\text{cos}(\mathbf{v}^{U}_{j},\mathbf{v}^{T}_{i})/\tau_{c})},
\end{split}
\end{align}
where in-batch negative samples are adopted, $N_{b}$ is the batch size (number of user sequences in a training batch), $\text{cos}(\cdot,\cdot)$ calculates the cosine similarity between two input vectors, and $\tau_{c}$ is the temperature for contrastive alignments. 

Our final training objective is a mixture of $\mathcal{L}_{NIG}$ and contrastive losses as shown in Fig.~\ref{fig:framework}:
\begin{align}
\begin{split}\label{formula:mixloss}
\mathcal{L}_{CALRec}=(1-\alpha-\beta)\mathcal{L}_{NIG}+\alpha\mathcal{L}_{TT}+\beta\mathcal{L}_{UT}.
\end{split}
\end{align}
\subsection{Quasi-Round-Robin BM25 Retrieval}
\label{s:method_bm25}
At inference, we prompt the model with the text description of user's history of interactions (see also \S\ref{s:method_template}). To obtain more than one candidate next-item prediction, we conduct temperature sampling~\cite{fan-etal-2018-hierarchical} 
 to generate $N_{\text{gen}}$ pieces of text as $N_{\text{gen}}$ predictions for the next item. Each of the $N_{\text{gen}}$ predictions is associated with its sequence score 
$s$ (i.e., log probability, which is a negative real number typically in the range of $(-10, 0)$, as observed on our data).

\begin{algorithm}[!th]
  \algsetup{linenosize=\tiny}
  \footnotesize
\DontPrintSemicolon 
\caption{\small Quasi Round Robin BM25 Selection} 
\label{alg:rbbm25}
\newcommand\mycommfont[1]{\footnotesize\ttfamily\textcolor{blue}{#1}}
\SetCommentSty{mycommfont}
\SetKwInOut{Input}{Input}
\SetKwInOut{Output}{Output}
\SetKwFunction{KwSort}{SortDescending} 
\newcommand{\KwSortNamed}[1]{\texttt{by}=#1} 
\SetKwFunction{KwBM}{BM25} 
\SetKwFunction{kwLinearScale}{LinearScale} 
\SetKwFunction{KwEmptyMat}{EmptyMatrix}
\SetKwFunction{KwRemoveDuplicates}{RemoveDuplicates}
\SetKwFunction{KwMaxPool}{MaxPool}
\SetKwFunction{KwArgSort}{ArgSortDescending}

\Input{
$samples$, output texts sampled from the LLM (size = $N_{\text{gen}}$);\\

$logprobs$, LLM sequence scores of $samples$ (size = $N_{\text{gen}}$);\\

$C$, corpus of candidate items (size = $N_c$);\\

$N_{\text{preds}}$, a hyperparameter that determines the number of generated\\ text samples used for BM25 matching;\\

$\epsilon$, a hyperparameter that controls the modulation of BM25 scores\\ with LLM sequence scores.
}

\Output{The sorted list of item predictions.}

\BlankLine 
\tcp{Sort $samples$ by score and remove duplicates}
\texttt{sorted$\_$samples}, \texttt{sorted$\_$logprobs} $\gets$ \linebreak \KwSort{(\texttt{samples}, \texttt{logprobs}), \KwSortNamed{\texttt{logprobs}}} \;

\texttt{sorted$\_$samples}, \texttt{sorted$\_$logprobs} $\gets$ \linebreak \KwRemoveDuplicates{(\texttt{sorted$\_$samples}, \texttt{sorted$\_$logprobs})}  \;

\tcp{Keep the top $N_{\text{preds}}$ samples}

\texttt{top$\_$samples} $\gets$ \texttt{sorted$\_$samples}$[:N_{\text{preds}}]$\;
\texttt{top$\_$logprobs} $\gets$ \texttt{sorted$\_$logprobs}$[:N_{\text{preds}}]$\;
\texttt{match$\_$scores} $\gets$ \KwEmptyMat{$N_c$, $N_{\text{preds}}$}\;
\For{$i \gets 1$ \textbf{to} $N_{\text{preds}}$}{
\tcp{Find BM25 scores between entire corpus and each sample}
    \texttt{bm25$\_$scores} $\gets$ \KwBM{\texttt{top$\_$samples}[i], C}\;
    \texttt{bm25$\_$scores$\_$scaled} $\gets$ \kwLinearScale{\texttt{bm25$\_$scores}, \text{axis=0}}\;
  \tcp{Modulate by sequence score} 
    \texttt{modulated$\_$bm25$\_$scores} $\gets$ $e^{\epsilon \cdot \texttt{top$\_$logprobs[i]}} \times \texttt{bm25$\_$scores$\_$scaled}$ \;
    \texttt{match$\_$scores[:,i]} $\gets$ \texttt{modulated$\_$bm25$\_$scores}\;
}
\texttt{candidate$\_$scores} $\gets$  \KwMaxPool{\texttt{match$\_$scores}, \text{axis=1}} \tcp*{size = N$_c$}
\Return \KwArgSort{\texttt{candidate$\_$scores}}
\end{algorithm}

We remove duplicate generations, sort them by sequence scores in descend order, and retain at most top $N_{\text{preds}}$ unique predictions. Then, in order to match these generated texts with items from our corpus, we rely on BM25 retrieval~\cite{robertson2009probabilistic} to calculate the matching scores between each text prediction and the whole item corpus (with size $N_{c}$): this will result in BM25 matching scores organized into a matrix of shape ($N_{c}$, $N_{\text{preds}}$). In order to rank the candidate items, we devise the following heuristics which aims to select the closest matches according to both the LLM sequence scores and BM25 scores in a quasi-round-robin fashion across the different $N_{\text{preds}}$ text generations. First, we linearly scale each column of the matrix respectively to the range $[0,1]$ to make BM25 scores derived from different LLM predictions comparable. We then penalize each prediction’s BM25 scores (each column) by multiplying the BM25 scores by a factor of $e^{\epsilon s}$, where $s$ is the sequence score associated with the text prediction, and $\epsilon$ is a small positive constant so that $e^{\epsilon s}$ is smaller than but very close to $1.0$. This is to further modulate the BM25 text-based retrieval scores by the LLM generative scores. After that, we max-pool 
along the $N_{\text{preds}}$ axis and finally derive modulated BM25 matching scores 
in a single vector of size $N_{c}$, with information from all $N_{\text{preds}}$ recommendations fused in it (we can then derive the ranking of the ground-truth target item in the item corpus for calculating evaluation metric scores, or return top-K recommendations for users), as presented in 
Algorithm~\ref{alg:rbbm25}\footnote{In our preliminary investigation, we found that \textbf{1)} max-pooling significantly outperforms min-pooling and mean-pooling and \textbf{2)} there is obvious performance drop if we remove modulation since only max-pooling combined with modulation can make sure that top-1 LLM outputs' top-1 recommendation is still the top-1 recommendation in the merged list. \textbf{3)} The overall results are not sensitive to the value of $\epsilon$ as long as $\epsilon$ is small.}. Here, we provide another view into the round-robin nature of our approach: given that the scaled BM25 score of each best-match item from the non-duplicate $N_{\text{preds}}$ texts generated is $1$, when they are further modulated (scaled down) using $e^{\epsilon s}$, the final retrieval algorithm roughly follows a round-robin fashion, where the top-1 BM25 retrieval result of each of the generated text are in turn grouped as the final top $N_{\text{preds}}$ retrieval results, when $\epsilon$ is close to $0.0$. Since there are ties, we name our method `quasi-' rather than `strict-' round-robin.

\section{Experimental Setup}
\label{s:expsetup}
\sparagraph{LLM Selection} Our work leverages PaLM-2~\cite{anil2023palm,tay2023ul}, a state-of-the-art LLM that is pretrained with a mixture of different objectives and exhibits exceptional generalization capabilities on a variety of language tasks. Among PaLM-2 model variants, we choose PaLM-2 XXS as the LLM backbone for our tasks, balancing model capability with inference latency.\footnote{Here, XXS is a `T-shirt size' notation of the model size. S-, M- and L-sized PaLM-2 models reported in~\citet{anil2023palm} are all larger than our XXS-sized model.} Larger PaLM-2 S does not show a significant advantage against XXS model when both are without fine-tuning (see \S\ref{s:expresults}). Besides, the scale of our dataset (significantly smaller than text corpora for pretraining LLMs) may lead to overfitting when fine-tuning larger models~\cite{journals/corr/KirkpatrickPRVD16,mitchell2022fast}. 

\begin{table*}[h!]
\begin{center}
\resizebox{0.95\linewidth}{!}{%
\begin{tabular}{ccccccccccc}
\toprule 
\rowcolor{Gray}
\multirow{1}{*}{} &\multicolumn{1}{c}{} &\multicolumn{1}{c}{\textsc{Before Dedup}} &\multicolumn{8}{c}{\textsc{After Dedup}} \\
\rowcolor{Gray}
\multirow{1}{*}{Dataset} &\multicolumn{1}{c}{\# of Users} &\multicolumn{1}{c}{\% of duplicate Purchases} &\multicolumn{1}{c}{\# of Items} &\multicolumn{1}{c}{Total Purchase} &\multicolumn{1}{c}{Items/User} &\multicolumn{1}{c}{Purchases/Item} &\multicolumn{1}{c}{Density} &\multicolumn{1}{c}{Words/Item} &\multicolumn{1}{c}{Word Vocab Size} &\multicolumn{1}{c}{Avg. Word Freq} \\
\cmidrule(lr){3-3} \cmidrule{4-11} 

\multirow{1}{*}{Scientific} &\multicolumn{1}{c}{9461} &\multicolumn{1}{c}{5.5\%} &\multicolumn{1}{c}{5282} &\multicolumn{1}{c}{66644} &\multicolumn{1}{c}{7.04} &\multicolumn{1}{c}{12.62} &\multicolumn{1}{c}{0.0013} &\multicolumn{1}{c}{22.82} &\multicolumn{1}{c}{19178} &\multicolumn{1}{c}{6.3} \\
\multirow{1}{*}{Instruments} &\multicolumn{1}{c}{25577} &\multicolumn{1}{c}{4.0\%} &\multicolumn{1}{c}{10599} &\multicolumn{1}{c}{214526}  &\multicolumn{1}{c}{8.39} &\multicolumn{1}{c}{20.24} &\multicolumn{1}{c}{0.00079} &\multicolumn{1}{c}{18.43} &\multicolumn{1}{c}{22256} &\multicolumn{1}{c}{8.8} \\
\multirow{1}{*}{Arts} &\multicolumn{1}{c}{47197} &\multicolumn{1}{c}{9.4\%} &\multicolumn{1}{c}{22828} &\multicolumn{1}{c}{411449}  &\multicolumn{1}{c}{8.72} &\multicolumn{1}{c}{18.02} &\multicolumn{1}{c}{0.00038} &\multicolumn{1}{c}{21.38} &\multicolumn{1}{c}{40342} &\multicolumn{1}{c}{12.1} \\
\multirow{1}{*}{Office} &\multicolumn{1}{c}{44736 (50\%)} &\multicolumn{1}{c}{6.3\%} &\multicolumn{1}{c}{27482} &\multicolumn{1}{c}{352151}  &\multicolumn{1}{c}{7.87} &\multicolumn{1}{c}{12.81} &\multicolumn{1}{c}{0.00029} &\multicolumn{1}{c}{21.74} &\multicolumn{1}{c}{56687} &\multicolumn{1}{c}{10.5} \\
\multirow{1}{*}{Games} &\multicolumn{1}{c}{50940} &\multicolumn{1}{c}{4.7\%} &\multicolumn{1}{c}{17383} &\multicolumn{1}{c}{457060}  &\multicolumn{1}{c}{8.97} &\multicolumn{1}{c}{26.29} &\multicolumn{1}{c}{0.00051} &\multicolumn{1}{c}{16.39} &\multicolumn{1}{c}{17087} &\multicolumn{1}{c}{16.7} \\
\multirow{1}{*}{Pet} &\multicolumn{1}{c}{43135 (20\%)} &\multicolumn{1}{c}{5.9\%} &\multicolumn{1}{c}{37712} &\multicolumn{1}{c}{380623} &\multicolumn{1}{c}{8.82} &\multicolumn{1}{c}{10.09} &\multicolumn{1}{c}{0.00023} &\multicolumn{1}{c}{19.87} &\multicolumn{1}{c}{47339} &\multicolumn{1}{c}{15.8} \\
\bottomrule
\end{tabular}
}
\caption{Statistics of Amazon Review dataset.}
\label{table:DataStats}
\end{center}
\end{table*}

\rparagraph{Dataset} Our experiments adopt the established and publicly available \href{https://cseweb.ucsd.edu/~jmcauley/datasets/amazon_v2/}{Amazon Review dataset 2018}~\cite{ni-etal-2019-justifying}. Specifically, we use the $5$-core subsets where each user/item is found in at least $5$ interactions. We choose four item attributes: \textit{title, category, brand}, and \textit{price}, to construct the flattened text description of each item, and we truncate the length of their corresponding descriptions to $25$, $15$, $15$, and $15$ words, respectively. Following~\citet{10.1145/3580305.3599519}, the same $14$ product categories are involved in our research and they are all used for Stage \RN{1} multi-category joint fine-tuning ($3.59$M users). Among them, $7$ categories ($3.37$M users) are only for Stage \RN{1}
training where their entire user purchase records are used\footnote{The $7$ categories are `Automotive', `Cell Phones and Accessories', `Clothing Shoes and Jewelry
', `Electronics', ` Grocery and Gourmet Food', `Home and Kitchen', and `Movies and TV'. The aim of leveraging large volume of data in Stage \RN{1} is to benefit other categories in Stage \RN{2} as a result of transfer learning.
}; one category `CD and Vinyls' is used for model selection purpose (29K users) during Stage \RN{1} training where the last item for each user is left for validation and only the first $n-1$ items are adopted for training. Finally, the remaining $6$ categories are adopted for category-specific fine-tuning (Stage \RN{2}) and evaluation ($0.22$M Users)\footnote{The $6$ categories are `Industrial and Scientific', `Musical Instruments', `Arts, Crafts and Sewing', `Office Products', `Video Games', `Pet Supplies'. For brevity, we also denote them as `Scientific', `Instruments', `Arts', `Office', `Games', and `Pet', respectively.}, where the first $n-2$ items are used for training in both Stages \RN{1} and \RN{2}, the penultimate item is used for validation in Stage \RN{2}, and the very last item for evaluation. We also conduct data deduplication (dedup) on user sequences, which we describe later in next paragraph. For large categories including `Office' and `Pet', we randomly sampled $50\%$ and $20\%$ users respectively. The detailed statistics of our data in Stage \RN{2} (after deduplication) are presented in Table~\ref{table:DataStats},
%
%
%
where `Density' is defined as $\text{Total Purchase}/(\# \text{of Users} \times \# \text{of Items})$, and the ‘Word Vocab Size’ denotes the lowercase natural word types over the item corpus spanning the four attributes used. 

\rparagraph{User Sequence Deduplication} We have found that the Amazon Review data include a non-negligible percentage of consecutive duplicate events in user sequences, where the duplication is identified as two consecutive activities by the same user that have the same item ID (ASIN number), review text, rating, and the same review timestamp. As indicated in Table~\ref{table:DataStats}, around $4\%$ to $9.4\%$ of purchases are exact duplicates of the previous ones in their user sequences. We conduct deduplication (dedup) on all user sequences following our strict definition of duplication above, while we still keep two consecutive purchases of the same item, if they occurred at different times (i.e., they come with different timestamps) or come with different ratings/review text since they are likely the users' real actions. We note that some previous work, e.g. ~\cite{10.1145/3580305.3599519}, did not conduct user sequence deduplication, which may lead a recommender system to learn a trivial strategy of always recommending the last item in the input sequence.

\rparagraph{Implementation Details} We train our CALRec based on XXS-sized PaLM-2 on a cluster of TPUv4 chips~\cite{2023arXiv230401433J}; our code is implemented with JAX~\cite{jax2018github}\footnote{\url{https://github.com/google/jax}}
and PAX\footnote{\url{https://github.com/google/paxml}}. 
We adopt a training batch size of $512$ users, a learning rate of $1$e$-4$, a maximum input length of $1,024$ tokens, and a maximum output decoding length of $80$ tokens. As a data augmentation trick, during training, we randomly truncate the last $k$ items ($k\in\{0, 1, 2, 3, 4\}$) from the full user record (with $n$ items) and always use the last item after truncation (i.e., the $(n-k)$-th item in the original user record) as the target item. We train CALRec with $\alpha=0.125,\beta=-0.025$ for up to $150,000$ steps in Stage \RN{1} and adopt $\tau_{c}=0.5$ (see Eqs.~\ref{formula:CL_TT} \&~\ref{formula:CL_UT}) as the configuration yields best validation scores. We adopt the checkpoint with $135,000$ training steps in Stage \RN{1} selected on the validation data category, and then further fine-tune the model for Stage \RN{2}. There are two exceptions: Scientific\footnote{We pick the Stage \RN{1} model at $45,000$ training steps since `Scientific' is significantly smaller than other categories and its own validation set indicates that overfitting occurs at $135,000$ steps and then continue with Stage \RN{2} training.}, and Games\footnote{For Games category, we adopt $\alpha=0.125,\beta=0$ for both stages and its Stage \RN{1} checkpoint at $105,000$ steps is selected using the Stage \RN{1} validation set. It is because we found on the `Games' own validation set, the new setup outperforms the previous (default) setup.}. 
 All our model selection is based on the NDCG@10 metric following previous work (specifically we adopt its pessimistic estimation which will be defined later in this section). 
 For each case, the same training objective is used in Stages \RN{1} and \RN{2}. 
 At inference, the temperature for decoding is $0.5$, $N_{gen}=32$, $N_{pred}=10$ and $\epsilon=1/5000$. There are two hyper-parameters inherent to BM25: we adopt the default values from the original BM25 API, $k_{1}=1.5$ and $b=0.75$, throughout all our experiments and did not further tune them since we found that the model performance is not sensitive to them. 

\rparagraph{Baselines} 
In this work, we consider a comprehensive set of baseline approaches. SASRec~\cite{8594844} and BERT4Rec~\cite{10.1145/3357384.3357895} are pure ID-Based approaches, Recformer~\cite{10.1145/3580305.3599519} and the `Last Item Repeater' (LIR) are purely text-based, and FDSA~\cite{ijcai2019p600} and S$^{3}$-Rec~\cite{10.1145/3340531.3411954} leverage both ID and text. UniSRec~\cite{10.1145/3534678.3539381} has two variants and we report its performance in both text-only mode and ID+text mode. More details about each baseline are provided in what follows.\footnote{In our experiments, LIR is implemented on our own, Recformer follows its official implementation. The rest baselines are all provided by the official code repo of UniSRec which is based on RecBole~\cite{zhao2021recbole}:
the attribute embeddings of FDSA and S$^{3}$-Rec are text features from our strongest baseline UniSRec.}
\begin{itemize}[labelwidth=0pt, labelindent=0pt, itemsep=0pt, topsep=0pt, leftmargin=*]
\item\textbf{SASRec}~\cite{8594844} is a causal sequential recommendation approach implemented with an encoder-only unidirectional transformer. It takes in a sequence of item IDs and predicts the ID embedding of the next item in the last position.
\item\textbf{BERT4Rec}~\cite{10.1145/3357384.3357895} adopts a bidirectional transformer architecture for sequential recommendation. Its input consists of a sequence of item IDs and it is trained with the masked language modeling objective. At inference, a ‘mask’ token is appended to the end of the sequence to predict the next item.
\item\textbf{FDSA}~\cite{ijcai2019p600} applies two self-attentional blocks to address ID sequence and attribute feature sequence, respectively, and fuses their final representations for next-item prediction. 
\item\textbf{S$^3$-Rec}~\cite{10.1145/3340531.3411954} maximizes the mutual information between ID and attribute embeddings via contrastive objectives and it is implemented with an encoder-only transformer network. It consists of a pretraining stage and a fine-tuning stage, both conducted on domain-specific data. 
\item\textbf{UniSRec}~\cite{10.1145/3534678.3539381} separately encodes each item’s text via a pretrained and fixed BERT cascaded by a trainable adapter. Given a sequence of item embeddings, a sequence representation is then modeled with another small transformer network. We adopt the officially released UniSRec model pretrained on multiple domains and fine-tune it on specific Amazon categories, respectively. We report two fine-tuning setups of UniSRec: (1) text-only setup, (2) ID+text setup where an item embedding is the sum of its ID and text representations.
\item\textbf{Recformer}~\cite{10.1145/3580305.3599519} relies on the encoder-only Longformer architecture~\cite{beltagy2020longformer} where token type and item position embeddings are proposed to facilitate learning item and user representations. It also features a pretraining stage on multiple categories with a combination of standard Masked Language Modeling (MLM) loss and a contrastive objective aligning user history and target item representations. We adopt Recformer’s officially released pretrained checkpoint and fine-tune it on specific categories, respectively.
\item\textbf{Last Item Repeater}, 
or \textbf{LIR} hereafter, denotes a simple yet effective baseline strategy which always recommends the last item in the input sequence as our training-free, text-statistic baseline. Based on the text of the last item, we rely on BM25 to rank the items in the full item corpus.
\end{itemize}

\rparagraph{Evaluation Metrics} Following standard practices from prior work, we report NDCG@10, Recall@K (K$\in\{1,10\}$) and MRR@$\infty$ scores. Different from previous work, we report both `optimistic' (\textit{opt}) and `pessimistic' (\textit{pes}) calculations of the metrics: when there is a tie such that the ground-truth target item and other candidate items have the same matching score, the ground-truth item ranks first in (\textit{opt}) and ranks last in (\textit{pes}) calculation, respectively. It is because \textbf{1)} there are distinct items (with different item IDs) with the same text description and \textbf{2)} in rare cases, we found that BM25 may yield identical scores for very similar text pieces. We still treat items in case \textbf{1)} as distinct items because in real production environments vendors/content creators may intentionally or unintentionally duplicate items. Of course, this will not cause any issue for ID-based approaches due to distinct item IDs, where (\textit{opt}) and (\textit{pes}) scores are the same. In the official implementations of our text-based baselines,~\citet{10.1145/3580305.3599519} only calculates (\textit{opt}) metric scores which may result in unfair comparisons with ID-based baselines. ~\citet{10.1145/3534678.3539381} applies ``\texttt{\href{https://pytorch.org/docs/stable/generated/torch.topk.html}{torch.topk()}}'' to derive exactly top $K$ item ids for calculating Recall/NDCG@K: if the $K$-th and the $K+1$-th items are having the same score, only one of them will be included in the output list. So its metric score is a number between our
(\textit{opt}) and (\textit{pes}) estimates.\footnote{To provide more insights on our (\textit{opt})/(\textit{pes}) metrics, readers could consider an extreme scenario where the model output is empty and thus when using BM25 retrieval the matching score is $0$ for all candidate items. In this case, the optimistic Recall@K scores will always be $1.0$, while the pessimistic scores will always be $0.0$. To provide a fair assessment of the model's ability to uniquely identify candidate items, we have thus decided to report both (\textit{opt}) and (\textit{pes}) scores.} All our evaluations are performed on user-sequence-deduplicated data. 

\section{Experimental Results}
\label{s:expresults}
\begin{table*}[h!]
\begin{center}
\resizebox{0.89\linewidth}{!}{%
\begin{tabular}{cccccccccccc}
\toprule 
\rowcolor{Gray}
\multirow{1}{*}{}
&\multicolumn{1}{c}{} &\multicolumn{2}{c}{\textsc{ID-Only Methods}} &\multicolumn{3}{c}{\textsc{ID+Text Methods}} &\multicolumn{4}{c}{\textsc{Text-Only Methods}} &\multicolumn{1}{c}{} \\
\rowcolor{Gray}
\multirow{1}{*}{Dataset} &\multicolumn{1}{c}{\makecell{  Metric  \\ \ \ (\textit{opt})/(\textit{pes}) \ \ }} &\multicolumn{1}{c}{BERT4Rec} &\multicolumn{1}{c}{SASRec} &\multicolumn{1}{c}{FDSA} &\multicolumn{1}{c}{S$^3$-Rec} &\multicolumn{1}{c}{UniSRec} &\multicolumn{1}{c}{UniSRec} &\multicolumn{1}{c}{RecFormer} &\multicolumn{1}{c}{LIR} &\multicolumn{1}{c}{CALRec} &\multicolumn{1}{c}{\makecell{Improv.\\(\textit{opt})}} \\

\cmidrule(lr){3-4} \cmidrule(lr){5-7} \cmidrule{8-11}
\multirow{4}{*}{Scientific}
&\multicolumn{1}{c}{NDCG@10} &\multicolumn{1}{c}{0.0411} &\multicolumn{1}{c}{0.0555} &\multicolumn{1}{c}{0.0594} &\multicolumn{1}{c}{0.0414} &\multicolumn{1}{c}{\cellcolor{blue!7}0.0644} &\multicolumn{1}{c}{0.0620/0.0580} &\multicolumn{1}{c}{0.0639/0.0588} &\multicolumn{1}{c}{0.0565/0.0486} &\multicolumn{1}{c}{\textbf{0.0788}/\underline{0.0740}} &\multicolumn{1}{c}{22.4\%}\\
&\multicolumn{1}{c}{Recall@1} &\multicolumn{1}{c}{0.0239} &\multicolumn{1}{c}{0.0079} &\multicolumn{1}{c}{\cellcolor{blue!7}0.0357} &\multicolumn{1}{c}{0.0172} &\multicolumn{1}{c}{0.0199} &\multicolumn{1}{c}{0.0242/0.0140} &\multicolumn{1}{c}{0.0298/0.0168} &\multicolumn{1}{c}{0.0142/0.0052} &\multicolumn{1}{c}{\textbf{0.0511}/\underline{0.0389}} &\multicolumn{1}{c}{43.1\%}\\
&\multicolumn{1}{c}{Recall@10} &\multicolumn{1}{c}{0.0636} &\multicolumn{1}{c}{0.1065} &\multicolumn{1}{c}{0.0887} &\multicolumn{1}{c}{0.0727} &\multicolumn{1}{c}{\cellcolor{blue!7}\textbf{\underline{0.1181}}} &\multicolumn{1}{c}{0.1101/0.1100} &\multicolumn{1}{c}{0.1058/0.1057} &\multicolumn{1}{c}{0.1005/0.0933} &\multicolumn{1}{c}{0.1124/0.1117} &\multicolumn{1}{c}{-4.8\%}\\
&\multicolumn{1}{c}{MRR} &\multicolumn{1}{c}{0.0389} &\multicolumn{1}{c}{0.0456} &\multicolumn{1}{c}{0.0555} &\multicolumn{1}{c}{0.0380} &\multicolumn{1}{c}{0.0555} &\multicolumn{1}{c}{0.0546/0.0492} &\multicolumn{1}{c}{\cellcolor{blue!7}0.0570/0.0502} &\multicolumn{1}{c}{0.0461/0.0385} &\multicolumn{1}{c}{\textbf{0.0730}/\underline{0.0667}} &\multicolumn{1}{c}{28.1\%}\\
\cmidrule(lr){3-4} \cmidrule(lr){5-7} \cmidrule{8-11}
\multirow{4}{*}{Instruments}
&\multicolumn{1}{c}{NDCG@10} &\multicolumn{1}{c}{0.0680} &\multicolumn{1}{c}{0.0623} &\multicolumn{1}{c}{\cellcolor{blue!7}0.0796} &\multicolumn{1}{c}{0.0606} &\multicolumn{1}{c}{0.0709} &\multicolumn{1}{c}{0.0661/0.0655} &\multicolumn{1}{c}{0.0596/0.0585} &\multicolumn{1}{c}{0.0344/0.0338} &\multicolumn{1}{c}{\textbf{0.0909}/\underline{0.0905}} &\multicolumn{1}{c}{14.2\%}\\
&\multicolumn{1}{c}{Recall@1} &\multicolumn{1}{c}{0.0494} &\multicolumn{1}{c}{0.0158} &\multicolumn{1}{c}{\cellcolor{blue!7}0.0574} &\multicolumn{1}{c}{0.0202} &\multicolumn{1}{c}{0.0237} &\multicolumn{1}{c}{0.0251/0.0236} &\multicolumn{1}{c}{0.0266/0.0239} &\multicolumn{1}{c}{0.0040/0.0031} &\multicolumn{1}{c}{\textbf{0.0718}/\underline{0.0706}} &\multicolumn{1}{c}{25.1\%}\\
&\multicolumn{1}{c}{Recall@10} &\multicolumn{1}{c}{0.0915} &\multicolumn{1}{c}{0.1130} &\multicolumn{1}{c}{0.1092} &\multicolumn{1}{c}{0.1048} &\multicolumn{1}{c}{\cellcolor{blue!7}\textbf{\underline{0.1255}}} &\multicolumn{1}{c}{0.1141/0.1141} &\multicolumn{1}{c}{0.0940/0.0940} &\multicolumn{1}{c}{0.0696/0.0692} &\multicolumn{1}{c}{0.1158/0.1158} &\multicolumn{1}{c}{-7.7\%}\\
&\multicolumn{1}{c}{MRR} &\multicolumn{1}{c}{0.0658} &\multicolumn{1}{c}{0.0528} &\multicolumn{1}{c}{\cellcolor{blue!7}0.0765} &\multicolumn{1}{c}{0.0526} &\multicolumn{1}{c}{0.0613} &\multicolumn{1}{c}{0.0576/0.0569} &\multicolumn{1}{c}{0.0535/0.0521} &\multicolumn{1}{c}{0.0263/0.0257} &\multicolumn{1}{c}{\textbf{0.0864}/\underline{0.0857}} &\multicolumn{1}{c}{12.9\%}\\
\cmidrule(lr){3-4} \cmidrule(lr){5-7} \cmidrule{8-11}
\multirow{4}{*}{Arts}
&\multicolumn{1}{c}{NDCG@10} &\multicolumn{1}{c}{0.0547} &\multicolumn{1}{c}{0.0619} &\multicolumn{1}{c}{0.0726} &\multicolumn{1}{c}{0.0602} &\multicolumn{1}{c}{\cellcolor{blue!7}0.0729} &\multicolumn{1}{c}{0.0630/0.0624} &\multicolumn{1}{c}{0.0662/0.0653} &\multicolumn{1}{c}{0.0443/0.0431} &\multicolumn{1}{c}{\textbf{0.0864}/\underline{0.0853}} &\multicolumn{1}{c}{18.5\%}\\
&\multicolumn{1}{c}{Recall@1} &\multicolumn{1}{c}{0.0347} &\multicolumn{1}{c}{0.0147} &\multicolumn{1}{c}{\cellcolor{blue!7}0.0460} &\multicolumn{1}{c}{0.0216} &\multicolumn{1}{c}{0.0213} &\multicolumn{1}{c}{0.0220/0.0208} &\multicolumn{1}{c}{0.0279/0.0254} &\multicolumn{1}{c}{0.0067/0.0051} &\multicolumn{1}{c}{\textbf{0.0636}/\underline{0.0610}} &\multicolumn{1}{c}{38.3\%}\\
&\multicolumn{1}{c}{Recall@10} &\multicolumn{1}{c}{0.0799} &\multicolumn{1}{c}{0.1118} &\multicolumn{1}{c}{0.1061} &\multicolumn{1}{c}{0.1052} &\multicolumn{1}{c}{\cellcolor{blue!7}\textbf{\underline{0.1320}}} &\multicolumn{1}{c}{0.1103/0.1102} &\multicolumn{1}{c}{0.1095/0.1095} &\multicolumn{1}{c}{0.0859/0.0850} &\multicolumn{1}{c}{0.1140/0.1139} &\multicolumn{1}{c}{-13.6\%}\\
&\multicolumn{1}{c}{MRR} &\multicolumn{1}{c}{0.0513} &\multicolumn{1}{c}{0.0519} &\multicolumn{1}{c}{\cellcolor{blue!7}0.0680} &\multicolumn{1}{c}{0.0520} &\multicolumn{1}{c}{0.0615} &\multicolumn{1}{c}{0.0546/0.0540} &\multicolumn{1}{c}{0.0582/0.0569} &\multicolumn{1}{c}{0.0344/0.0331} &\multicolumn{1}{c}{\textbf{0.0815}/\underline{0.0801}} &\multicolumn{1}{c}{19.9\%}\\
\cmidrule(lr){3-4} \cmidrule(lr){5-7} \cmidrule{8-11}
\multirow{4}{*}{Office}
&\multicolumn{1}{c}{NDCG@10} &\multicolumn{1}{c}{0.0545} &\multicolumn{1}{c}{0.0602} &\multicolumn{1}{c}{\cellcolor{blue!7}0.0749} &\multicolumn{1}{c}{0.0607} &\multicolumn{1}{c}{0.0713} &\multicolumn{1}{c}{0.0625/0.0616} &\multicolumn{1}{c}{0.0687/0.0676} &\multicolumn{1}{c}{0.0391/0.0379} &\multicolumn{1}{c}{\textbf{0.0976}/\underline{0.0966}} &\multicolumn{1}{c}{30.3\%}\\
&\multicolumn{1}{c}{Recall@1} &\multicolumn{1}{c}{0.0403} &\multicolumn{1}{c}{0.0134} &\multicolumn{1}{c}{\cellcolor{blue!7}0.0547} &\multicolumn{1}{c}{0.0291} &\multicolumn{1}{c}{0.0213} &\multicolumn{1}{c}{0.0250/0.0228} &\multicolumn{1}{c}{0.0328/0.0302} &\multicolumn{1}{c}{0.0056/0.0040} &\multicolumn{1}{c}{\textbf{0.0761}/\underline{0.0734}} &\multicolumn{1}{c}{39.1\%}\\
&\multicolumn{1}{c}{Recall@10} &\multicolumn{1}{c}{0.0709} &\multicolumn{1}{c}{0.1033} &\multicolumn{1}{c}{0.0982} &\multicolumn{1}{c}{0.0941} &\multicolumn{1}{c}{\cellcolor{blue!7}\textbf{\underline{0.1220}}} &\multicolumn{1}{c}{0.1023/0.1022} &\multicolumn{1}{c}{0.1063/0.1063} &\multicolumn{1}{c}{0.0735/0.0728} &\multicolumn{1}{c}{0.1213/0.1213} &\multicolumn{1}{c}{-0.6\%}\\
&\multicolumn{1}{c}{MRR} &\multicolumn{1}{c}{0.0520} &\multicolumn{1}{c}{0.0499} &\multicolumn{1}{c}{\cellcolor{blue!7}0.0710} &\multicolumn{1}{c}{0.0535} &\multicolumn{1}{c}{0.0597} &\multicolumn{1}{c}{0.0545/0.0533} &\multicolumn{1}{c}{0.0603/0.0588} &\multicolumn{1}{c}{0.0309/0.0296} &\multicolumn{1}{c}{\textbf{0.0925}/\underline{0.0911}} &\multicolumn{1}{c}{30.3\%}\\
\cmidrule(lr){3-4} \cmidrule(lr){5-7} \cmidrule{8-11}
\multirow{4}{*}{Games}
&\multicolumn{1}{c}{NDCG@10} &\multicolumn{1}{c}{0.0334} &\multicolumn{1}{c}{0.0451} &\multicolumn{1}{c}{\cellcolor{blue!7}0.0526} &\multicolumn{1}{c}{0.0419} &\multicolumn{1}{c}{0.0501} &\multicolumn{1}{c}{0.0408/0.0407} &\multicolumn{1}{c}{0.0377/0.0360} &\multicolumn{1}{c}{0.0211/0.0197} &\multicolumn{1}{c}{\textbf{0.0595}/\underline{0.0585}} &\multicolumn{1}{c}{13.1\%}\\
&\multicolumn{1}{c}{Recall@1} &\multicolumn{1}{c}{0.0112} &\multicolumn{1}{c}{0.0039} &\multicolumn{1}{c}{\cellcolor{blue!7}0.0210} &\multicolumn{1}{c}{0.0080} &\multicolumn{1}{c}{0.0075} &\multicolumn{1}{c}{0.0111/0.0107} &\multicolumn{1}{c}{0.0126/0.0091} &\multicolumn{1}{c}{0.0015/0.0009} &\multicolumn{1}{c}{\textbf{0.0295}/\underline{0.0277}} &\multicolumn{1}{c}{40.5\%}\\
&\multicolumn{1}{c}{Recall@10} &\multicolumn{1}{c}{0.0646} &\multicolumn{1}{c}{0.0989} &\multicolumn{1}{c}{0.0957} &\multicolumn{1}{c}{0.0884} &\multicolumn{1}{c}{\cellcolor{blue!7}\textbf{\underline{0.1074}}} &\multicolumn{1}{c}{0.0820/0.0820} &\multicolumn{1}{c}{0.0724/0.0723} &\multicolumn{1}{c}{0.0442/0.0427} &\multicolumn{1}{c}{0.0986/0.0984} &\multicolumn{1}{c}{-8.2\%}\\
&\multicolumn{1}{c}{MRR} &\multicolumn{1}{c}{0.0313} &\multicolumn{1}{c}{0.0380} &\multicolumn{1}{c}{\cellcolor{blue!7}0.0483} &\multicolumn{1}{c}{0.0367} &\multicolumn{1}{c}{0.0427} &\multicolumn{1}{c}{0.0372/0.0370} &\multicolumn{1}{c}{0.0349/0.0326} &\multicolumn{1}{c}{0.0171/0.0159} &\multicolumn{1}{c}{\textbf{0.0510}/\underline{0.0498}} &\multicolumn{1}{c}{5.6\%}\\
\cmidrule(lr){3-4} \cmidrule(lr){5-7} \cmidrule{8-11}
\multirow{4}{*}{Pet}
&\multicolumn{1}{c}{NDCG@10} &\multicolumn{1}{c}{0.0376} &\multicolumn{1}{c}{0.0448} &\multicolumn{1}{c}{0.0537} &\multicolumn{1}{c}{0.0392} &\multicolumn{1}{c}{0.0574} &\multicolumn{1}{c}{0.0539/0.0464} &\multicolumn{1}{c}{\cellcolor{blue!7}0.0590/0.0471} &\multicolumn{1}{c}{0.0391/0.0354} &\multicolumn{1}{c}{\textbf{0.0736}/\underline{0.0693}} &\multicolumn{1}{c}{24.8\%}\\
&\multicolumn{1}{c}{Recall@1} &\multicolumn{1}{c}{0.0276} &\multicolumn{1}{c}{0.0093} &\multicolumn{1}{c}{\cellcolor{blue!7}0.0394} &\multicolumn{1}{c}{0.0162} &\multicolumn{1}{c}{0.0202} &\multicolumn{1}{c}{0.0295/0.0129} &\multicolumn{1}{c}{0.0377/0.0130} &\multicolumn{1}{c}{0.0094/0.0030} &\multicolumn{1}{c}{\textbf{0.0570}/\underline{0.0465}} &\multicolumn{1}{c}{44.7\%}\\
&\multicolumn{1}{c}{Recall@10} &\multicolumn{1}{c}{0.0499} &\multicolumn{1}{c}{0.0773} &\multicolumn{1}{c}{0.0710} &\multicolumn{1}{c}{0.0647} &\multicolumn{1}{c}{\cellcolor{blue!7}\textbf{\underline{0.0970}}} &\multicolumn{1}{c}{0.0829/0.0826} &\multicolumn{1}{c}{0.0826/0.0818} &\multicolumn{1}{c}{0.0680/0.0673} &\multicolumn{1}{c}{0.0937/0.0934} &\multicolumn{1}{c}{-3.4\%}\\
&\multicolumn{1}{c}{MRR} &\multicolumn{1}{c}{0.0365} &\multicolumn{1}{c}{0.0382} &\multicolumn{1}{c}{0.0522} &\multicolumn{1}{c}{0.0349} &\multicolumn{1}{c}{0.0503} &\multicolumn{1}{c}{0.0498/0.0399} &\multicolumn{1}{c}{\cellcolor{blue!7}0.0549/0.0393} &\multicolumn{1}{c}{0.0317/0.0272} &\multicolumn{1}{c}{\textbf{0.0696}/\underline{0.0639}} &\multicolumn{1}{c}{26.8\%}\\
\cmidrule(lr){3-4} \cmidrule(lr){5-7} \cmidrule{8-11}
\multirow{4}{*}{\makecell{\textit{Average}\\($6$ Categories)}}
&\multicolumn{1}{c}{NDCG@10} &\multicolumn{1}{c}{0.0482} &\multicolumn{1}{c}{0.0550} &\multicolumn{1}{c}{\cellcolor{blue!7}0.0655} &\multicolumn{1}{c}{0.0507} &\multicolumn{1}{c}{0.0645} &\multicolumn{1}{c}{0.0581/0.0558} &\multicolumn{1}{c}{0.0592/0.0556} &\multicolumn{1}{c}{0.0391/0.0364} &\multicolumn{1}{c}{\textbf{0.0811}/\underline{0.0790}} &\multicolumn{1}{c}{23.8\%}\\
&\multicolumn{1}{c}{Recall@1} &\multicolumn{1}{c}{0.0312} &\multicolumn{1}{c}{0.0108} &\multicolumn{1}{c}{\cellcolor{blue!7}0.0424} &\multicolumn{1}{c}{0.0187} &\multicolumn{1}{c}{0.0190} &\multicolumn{1}{c}{0.0228/0.0175} &\multicolumn{1}{c}{0.0279/0.0197} &\multicolumn{1}{c}{0.0069/0.0036} &\multicolumn{1}{c}{\textbf{0.0582}/\underline{0.0530}} &\multicolumn{1}{c}{37.3\%}\\
&\multicolumn{1}{c}{Recall@10} &\multicolumn{1}{c}{0.0701} &\multicolumn{1}{c}{0.1018} &\multicolumn{1}{c}{0.0948} &\multicolumn{1}{c}{0.0883} &\multicolumn{1}{c}{\cellcolor{blue!7}\textbf{\underline{0.1170}}} &\multicolumn{1}{c}{0.1003/0.1002} &\multicolumn{1}{c}{0.0951/0.0949} &\multicolumn{1}{c}{0.0736/0.0717} &\multicolumn{1}{c}{0.1093/0.1091} &\multicolumn{1}{c}{-6.6\%}\\
&\multicolumn{1}{c}{MRR} &\multicolumn{1}{c}{0.0460} &\multicolumn{1}{c}{0.0461} &\multicolumn{1}{c}{\cellcolor{blue!7}0.0619} &\multicolumn{1}{c}{0.0446} &\multicolumn{1}{c}{0.0552} &\multicolumn{1}{c}{0.0514/0.0484} &\multicolumn{1}{c}{0.0531/0.0483} &\multicolumn{1}{c}{0.0311/0.0283} &\multicolumn{1}{c}{\textbf{0.0757}/\underline{0.0729}} &\multicolumn{1}{c}{22.3\%}\\

\bottomrule
\end{tabular}
}
\caption{Evaluation results of CALRec in comparison with different ID/Text-based sequential recommendation baselines (cf. \S\ref{s:expsetup}). For text-only methods, we report both optimisitc (\textit{opt}) and pessimistic (\textit{pes}) as `(\textit{opt})/(\textit{pes})'. For ID-based methods, (\textit{opt}) and (\textit{pes}) scores are the same. \textbf{\large Bold}: the highest  (\textit{opt}) scores. \underline{Underline}: the strongest (\textit{pes}) scores. We also report the relative improvement (\textit{opt}) comparing CALRec against (\textit{opt}) baseline SotA marked with \colorbox{blue!7}{\footnotesize \lstinline|Blue Background|} per data category per metric. Evaluation metrics are computed using the Amazon Review dataset after user sequence deduplication (see also \S\ref{s:expsetup}).}
\label{table:MainRes}
\end{center}
\end{table*}
\begin{table*}[h!]
\begin{center}
\resizebox{0.92\linewidth}{!}{%
\begin{tabular}{ccccccccccccc}
\toprule 
\rowcolor{Gray}
\multirow{1}{*}{}  &\multicolumn{4}{c}{\textsc{Scientific}} &\multicolumn{4}{c}{\textsc{Instruments}} &\multicolumn{4}{c}{\textsc{Games}}\\
\rowcolor{Gray}
\multirow{1}{*}{Model Variant}  &\multicolumn{1}{c}{NDCG@10} &\multicolumn{1}{c}{Recall@1} &\multicolumn{1}{c}{Recall@10} &\multicolumn{1}{c}{MRR} &\multicolumn{1}{c}{NDCG@10} &\multicolumn{1}{c}{Recall@1} &\multicolumn{1}{c}{Recall@10} &\multicolumn{1}{c}{MRR}  &\multicolumn{1}{c}{NDCG@10} &\multicolumn{1}{c}{Recall@1} &\multicolumn{1}{c}{Recall@10} &\multicolumn{1}{c}{MRR}\\

\cmidrule(lr){2-5} \cmidrule(lr){6-9} \cmidrule{10-13}

\multirow{1}{*}{CALRec} &\multicolumn{1}{c}{\makecell{\textbf{0.0788} /\\ \underline{0.0740}}} &\multicolumn{1}{c}{\makecell{\textbf{0.0511} /\\ \underline{0.0389}}} &\multicolumn{1}{c}{\makecell{\textbf{0.1124} /\\ \underline{0.1117}}} &\multicolumn{1}{c}{\makecell{\textbf{0.0730} /\\ \underline{0.0667}}} &\multicolumn{1}{c}{\makecell{\textbf{0.0909} /\\ \underline{0.0905}}} &\multicolumn{1}{c}{\makecell{\textbf{0.0718} /\\ \underline{0.0706}}} &\multicolumn{1}{c}{\makecell{\textbf{0.1158} /\\ \underline{0.1158}}} &\multicolumn{1}{c}{\makecell{\textbf{0.0864} /\\ \underline{0.0857}}} &\multicolumn{1}{c}{\makecell{\textbf{0.0595} /\\ \underline{0.0585}}} &\multicolumn{1}{c}{\makecell{\textbf{0.0295} /\\ \underline{0.0277}}} &\multicolumn{1}{c}{\makecell{0.0986 /\\ 0.0984}} &\multicolumn{1}{c}{\makecell{\textbf{0.0510} /\\ \underline{0.0498}}} \\
\cmidrule(lr){2-5} \cmidrule(lr){6-9} \cmidrule{10-13}
\multirow{1}{*}{w/o Multi-Category Joint Fine-Tuning (A)} &\multicolumn{1}{c}{\makecell{0.0727 /\\ 0.0675}} &\multicolumn{1}{c}{\makecell{0.04408 /\\ 0.0314}} &\multicolumn{1}{c}{\makecell{0.1070 /\\ 0.1061}} &\multicolumn{1}{c}{\makecell{0.0670 /\\ 0.0602}} &\multicolumn{1}{c}{\makecell{0.0870 /\\ 0.0862}} &\multicolumn{1}{c}{\makecell{0.0673 /\\ 0.0658}} &\multicolumn{1}{c}{\makecell{0.1119 /\\ 0.1115}} &\multicolumn{1}{c}{\makecell{0.0822 /\\ 0.0812}} &\multicolumn{1}{c}{\makecell{0.0583 /\\ 0.0570}} &\multicolumn{1}{c}{\makecell{0.0286 /\\ 0.0265}} &\multicolumn{1}{c}{\makecell{0.0966 /\\ 0.0959}} &\multicolumn{1}{c}{\makecell{0.0501 /\\ 0.0486}} \\
\cmidrule(lr){2-5} \cmidrule(lr){6-9} \cmidrule{10-13}
\multirow{1}{*}{w/o Category-Specific Fine-Tuning (B)} &\multicolumn{1}{c}{\makecell{0.0717 /\\ 0.0670}} &\multicolumn{1}{c}{\makecell{0.0424 /\\ 0.0300}} &\multicolumn{1}{c}{\makecell{0.1077 /\\ 0.1073}} &\multicolumn{1}{c}{\makecell{0.0653 /\\ 0.0588}} &\multicolumn{1}{c}{\makecell{0.0867 /\\ 0.0862}} &\multicolumn{1}{c}{\makecell{0.0715 /\\ 0.0702}} &\multicolumn{1}{c}{\makecell{0.1048 /\\ 0.1048}} &\multicolumn{1}{c}{\makecell{0.0840 /\\ 0.0834}} &\multicolumn{1}{c}{\makecell{0.0539 /\\ 0.0530}} &\multicolumn{1}{c}{\makecell{0.0257 /\\ 0.0238}} &\multicolumn{1}{c}{\makecell{0.0904 /\\ 0.0902}} &\multicolumn{1}{c}{\makecell{0.0463 /\\ 0.0451}} \\
\cmidrule(lr){2-5} \cmidrule(lr){6-9} \cmidrule{10-13}
\multirow{1}{*}{w/o Contrastive Alignment} &\multicolumn{1}{c}{\makecell{0.0771 /\\ 0.0724}} &\multicolumn{1}{c}{\makecell{0.0495 /\\ 0.0376}} &\multicolumn{1}{c}{\makecell{0.1109 /\\ 0.1105}} &\multicolumn{1}{c}{\makecell{0.0712 /\\ 0.0649}} &\multicolumn{1}{c}{\makecell{0.0905 /\\ 0.0899}} &\multicolumn{1}{c}{\makecell{0.0711 /\\ 0.0696}} &\multicolumn{1}{c}{\makecell{0.1154 /\\ 0.1153}} &\multicolumn{1}{c}{\makecell{0.0858 /\\ 0.0850}} &\multicolumn{1}{c}{\makecell{0.0593 /\\ 0.0584}} &\multicolumn{1}{c}{\makecell{0.0289 /\\ 0.0272}} &\multicolumn{1}{c}{\makecell{\textbf{0.0987} /\\ \underline{0.0985}}} &\multicolumn{1}{c}{\makecell{0.0507 /\\ 0.0495}} \\
\cmidrule(lr){2-5} \cmidrule(lr){6-9} \cmidrule{10-13}
\multirow{1}{*}{Pretrained PaLM-2 (w/o Fine-Tuning)} &\multicolumn{1}{c}{\makecell{0.0326 /\\ 0.0291}} &\multicolumn{1}{c}{\makecell{0.0090 /\\ 0.0055}} &\multicolumn{1}{c}{\makecell{0.0661 /\\ 0.0623}} &\multicolumn{1}{c}{\makecell{0.0267 /\\ 0.0234}} &\multicolumn{1}{c}{\makecell{0.0257 /\\ 0.0254}} &\multicolumn{1}{c}{\makecell{0.0041 /\\ 0.0037}} &\multicolumn{1}{c}{\makecell{0.0585 /\\ 0.0584}} &\multicolumn{1}{c}{\makecell{0.0197 /\\ 0.0193}} &\multicolumn{1}{c}{\makecell{0.0150 /\\ 0.0141}} &\multicolumn{1}{c}{\makecell{0.0031 /\\ 0.0024}} &\multicolumn{1}{c}{\makecell{0.0325 /\\ 0.0314}} &\multicolumn{1}{c}{\makecell{0.0134 /\\ 0.0127}} \\
\cmidrule(lr){2-5} \cmidrule(lr){6-9} \cmidrule{10-13}

\bottomrule
\end{tabular}
}
\caption{Ablation result showing the benefits of multi-stage fine-tuning and constrastive alignment to improve the performance of CALRec. \textbf{\large Bold}: the highest optimistic (\textit{opt}) scores. \underline{Underline}: the strongest pessimistic (\textit{pes}) scores.}
\label{table:ablation}
\end{center}
\end{table*}

\begin{table}[h!]
\begin{center}
\resizebox{0.999\linewidth}{!}{%
\begin{tabular}{ccccc}
\toprule 
\rowcolor{Gray}

\multirow{1}{*}{Metric: (\textit{opt})/(\textit{pes})} &\multicolumn{1}{c}{w/ Contr.} &\multicolumn{1}{c}{w/o Contr.} &\multicolumn{1}{c}{Improv.} &\multicolumn{1}{c}{$p$-value} \\


\multirow{1}{*}{NDCG@10} &\multicolumn{1}{c}{0.0811/0.0790} &\multicolumn{1}{c}{0.0803/0.0782} &\multicolumn{1}{c}{1.00\%/1.02\%} &\multicolumn{1}{c}{0.0076/0.0064} \\
\multirow{1}{*}{Recall@1} &\multicolumn{1}{c}{0.0582/0.0530} &\multicolumn{1}{c}{0.0574/0.0521} &\multicolumn{1}{c}{1.39\%/1.73\%} &\multicolumn{1}{c}{0.0027/0.0004} \\
\multirow{1}{*}{Recall@10} &\multicolumn{1}{c}{0.1093/0.1091} &\multicolumn{1}{c}{0.1084/0.1082} &\multicolumn{1}{c}{0.83\%/0.83\%} &\multicolumn{1}{c}{0.0178/0.019} \\
\multirow{1}{*}{MRR} &\multicolumn{1}{c}{0.0757/0.0729} &\multicolumn{1}{c}{0.0748/0.0720} &\multicolumn{1}{c}{1.20\%/1.25\%} &\multicolumn{1}{c}{0.0066/0.0037} \\

\bottomrule
\end{tabular}
}
\caption{Average scores (\textit{opt})/(\textit{pes}) on all $6$ data categories:  w/ contrastive alignment vs. w/o contrastive alignment.}
\label{table:ablation2}
\end{center}
\end{table}

\begin{table}[h!]
\begin{center}
\resizebox{0.96\linewidth}{!}{%
\begin{tabular}{ccccc}
\toprule 
\rowcolor{Gray}
\multirow{1}{*}{} &\multicolumn{1}{c}{} &\multicolumn{3}{c}{\textsc{Model Size}} \\
\rowcolor{Gray}
\multirow{1}{*}{Dataset} &\multicolumn{1}{c}{\makecell{  Metric  \\ \ \ (\textit{opt})/(\textit{pes}) \ \ }} &\multicolumn{1}{c}{XXXS} &\multicolumn{1}{c}{XXS} &\multicolumn{1}{c}{S} \\

\cmidrule(lr){3-5}
\multirow{4}{*}{Scientific}
&\multicolumn{1}{c}{NDCG@10} &\multicolumn{1}{c}{0.0299/0.0252} &\multicolumn{1}{c}{\textbf{0.0326}/0.0291} &\multicolumn{1}{c}{0.0321/\underline{0.0306}} \\
&\multicolumn{1}{c}{Recall@1} &\multicolumn{1}{c}{0.0089/0.0034} &\multicolumn{1}{c}{\textbf{0.0090}/0.0055} &\multicolumn{1}{c}{0.0082/\underline{0.0058}} \\
&\multicolumn{1}{c}{Recall@10} &\multicolumn{1}{c}{0.0592/0.0534} &\multicolumn{1}{c}{\textbf{0.0661}/0.0623} &\multicolumn{1}{c}{0.0660/\underline{0.0647}} \\
&\multicolumn{1}{c}{MRR} &\multicolumn{1}{c}{0.0246/0.0204} &\multicolumn{1}{c}{\textbf{0.0267}/0.0234} &\multicolumn{1}{c}{0.0262/\underline{0.0247}} \\
\cmidrule(lr){3-5}
\multirow{4}{*}{Instruments}
&\multicolumn{1}{c}{NDCG@10} &\multicolumn{1}{c}{0.0215/0.0210} &\multicolumn{1}{c}{\textbf{0.0257}/\underline{0.0254}} &\multicolumn{1}{c}{0.0234/0.0231} \\
&\multicolumn{1}{c}{Recall@1} &\multicolumn{1}{c}{0.0029/0.0022} &\multicolumn{1}{c}{0.0041/0.0037} &\multicolumn{1}{c}{\textbf{0.0043}/\underline{0.0039}} \\
&\multicolumn{1}{c}{Recall@10} &\multicolumn{1}{c}{0.0480/0.0476} &\multicolumn{1}{c}{\textbf{0.0585}/\underline{0.0584}} &\multicolumn{1}{c}{0.0512/0.0511} \\
&\multicolumn{1}{c}{MRR} &\multicolumn{1}{c}{0.0162/0.0156} &\multicolumn{1}{c}{\textbf{0.0197}/\underline{0.0193}} &\multicolumn{1}{c}{0.0188/0.0185} \\

\bottomrule
\end{tabular}
}
\caption{Performance results of pretrained PaLM-2 models of three distinct sizes without any fine-tuning for Sequential Recommendation. \textbf{\large Bold}: the highest optimistic (\textit{opt}) scores. \underline{Underline}: the strongest pessimistic (\textit{pes}) scores.}
\label{table:rawulm}
\end{center}
\end{table}

\begin{table}[h!]
\begin{center}
\resizebox{0.62\linewidth}{!}{%
\begin{tabular}{ccc}
\toprule 
\rowcolor{Gray}
\multirow{1}{*}{Dataset} &\multicolumn{1}{c}{Metric} &\multicolumn{1}{c}{(\textit{opt})/(\textit{pes})} \\
\cmidrule(lr){3-3}
\multirow{4}{*}{Scientific}
&\multicolumn{1}{c}{NDCG@10} &\multicolumn{1}{c}{0.0265/0.0232} \\
&\multicolumn{1}{c}{Recall@1} &\multicolumn{1}{c}{0.0051/0.0030}  \\
&\multicolumn{1}{c}{Recall@10} &\multicolumn{1}{c}{0.0582/0.0537} \\
&\multicolumn{1}{c}{MRR} &\multicolumn{1}{c}{0.0217/0.0189} \\
\cmidrule(lr){3-3}
\multirow{4}{*}{Instruments}
&\multicolumn{1}{c}{NDCG@10} &\multicolumn{1}{c}{0.0240/0.0237} \\
&\multicolumn{1}{c}{Recall@1} &\multicolumn{1}{c}{0.0028/0.0025}  \\
&\multicolumn{1}{c}{Recall@10} &\multicolumn{1}{c}{0.0584/ 0.0583} \\
&\multicolumn{1}{c}{MRR} &\multicolumn{1}{c}{0.0179/0.0175} \\
\bottomrule
\end{tabular}
}
\caption{Performance results of GPT4Rec's template using pretrained PaLM-2 XXS without any fine-tuning for Sequential Recommendation.}
\label{table:gpt4rectemplate}
\end{center}
\end{table}

\sparagraph{Main Results} Table~\ref{table:MainRes} demonstrates that the sequential recommendation capabilities of the proposed CALRec framework are stronger than our baselines. On the $6$ Amazon Review test categories, our model outperforms all $8$ baselines significantly in terms of NDCG@10, Recall@1 and MRR. In terms of Recall@10, our approach outperforms all \textsc{Text-Only} and \textsc{ID-Only} models across the board by a large margin (including the \textsc{Text-Only} variant of UniSRec), but slightly underperforms our strongest baseline, the \textsc{ID+Text} variant of UniSRec. For our method and $3$ \textsc{Text-Only} baselines, the results in Table~\ref{table:MainRes} verified that there is a gap between optimistic and pessimistic metric scores, demonstrating the necessity of reporting both. Meanwhile, it is worth noticing that even our pessimistic scores clearly surpass the optimistic scores of the $3$ \textsc{Text-Only} baseline models on all the $6$ data categories. In addition to the absolute scores, we also calculate the percentage gains comparing CALRec's optimistic scores against the baseline SotA for each data category and each evaluation metric (different method for different data category and metric). The results are presented in the last column of Table~\ref{table:MainRes}. We found that, CALRec is especially strong in Recall@1, outperforming baseline SotA by $37.3\%$ on average; CALRec outperforms baseline SotA by $23.8\%$ and $22.3\%$ on average in terms of NDCG@10 and MRR, respectively.

\rparagraph{Ablation Study} In Table~\ref{table:ablation}, we demonstrate the effectiveness of CALRec's key components via extensive ablation study on $3$ data categories (the results on the other $3$ categories show the same trends so we skip them due to space constraints).

\emph{Two-Stage Fine-Tuning:} First, it is obvious that our Stage \RN{1} multi-category joint fine-tuning and Stage \RN{2} category-specific fine-tuning are both effective, without which there will be an obvious drop in performance. Furthermore, it is worth mentioning that our model derived only with Stage \RN{1} is able to recommend items in all $6$ categories: this is especially valuable and easy to deploy in real-world applications, due to its multi-tenancy nature. Although the multi-category trained model (variant B in Table~\ref{table:ablation}) lags behind the full CALRec derived with both stages, it is already on par with variant A models (cf. Table~\ref{table:ablation}). 
We also compare CALRec and pretrained `off-the-shelf' PaLM-2 XXS in Table~\ref{table:ablation}; the results show that without any fine-tuning, PaLM-2 demonstrates poor performance on our sequential recommendation benchmarks.

\emph{Contrastive Objectives:} In Table~\ref{table:ablation2} we calculate the average scores on all $6$ categories for the setups with and without the auxiliary contrastive losses. Basically, the auxiliary contrastive objective can result in circa $0.8\%-1.7\%$ gains. We also conduct \href{https://en.wikipedia.org/wiki/Student\%27s_t-test#Dependent_t-test_for_paired_samples}{dependent paired sample t-test}~\cite{Ross2017} 
and report the $p$-values in the last column of Table~\ref{table:ablation2}: the gains are all statistically significant at $0.05$ level. 

\emph{Model Size:} We are also interested in how pretrained PaLM-2 \textit{without} any fine-tuning on our sequential recommendation data performs in our tasks and we present results with three PaLM-2 models of different model sizes (i.e., XXXS, XXS, and S) on `Scientific' and `Instruments' categories in Table~\ref{table:rawulm}. The input/output format is still the same: input data is from a single user. The results show that XXS and S models actually achieve similar performance and they both outperform the XXXS model. 

\emph{Template Design:} We conducted additional experiments on prompting the raw PaLM2 XXS model for recommendation adopting another prompt from GPT4Rec~\cite{li2023gpt4rec}, with results reported in Table~\ref{table:gpt4rectemplate}. Unlike our prompt template, their template doesn't include a shared per-item prefix and user suffix but leverages a one-off user suffix. Compared with our results reported in Table~\ref{table:rawulm} with the same raw PaLM-2 XXS, the new results are universally lower than ours (especially recall@1, relatively $30\%\sim45\%$ drop in scores), showing the effectiveness of our template design.

\section{Further Analyses and Discussion}
\label{s:discussion}
\begin{table*}[!ht]
\begin{center}
\resizebox{0.75\linewidth}{!}{%
\begin{tabular}{ccccccccc}
\toprule 
\rowcolor{Gray}
\multicolumn{1}{c}{} &\multicolumn{4}{c}{\textsc{w/o Dedup} - \textsc{Metric}: (\textit{opt})} &\multicolumn{4}{c}{\textsc{w/ Dedup} - \textsc{Metric}: (\textit{opt})} \\
\rowcolor{Gray}
\multirow{1}{*}{Dataset} &\multicolumn{1}{c}{NDCG@10} &\multicolumn{1}{c}{Recall@1} &\multicolumn{1}{c}{Recall@10} &\multicolumn{1}{c}{MRR} &\multicolumn{1}{c}{NDCG@10} &\multicolumn{1}{c}{Recall@1} &\multicolumn{1}{c}{Recall@10} &\multicolumn{1}{c}{MRR} \\
\cmidrule(lr){2-5} \cmidrule(lr){6-9}

\multirow{1}{*}{Scientific} &\multicolumn{1}{c}{0.088} &\multicolumn{1}{c}{0.0355} &\multicolumn{1}{c}{0.1381} &\multicolumn{1}{c}{0.0745} &\multicolumn{1}{c}{0.0565 (-36\%)} &\multicolumn{1}{c}{0.0142 (-60\%)} &\multicolumn{1}{c}{0.1005 (-27\%)} &\multicolumn{1}{c}{0.0461 (-38\%)} \\
\multirow{1}{*}{Instruments} &\multicolumn{1}{c}{0.0545} &\multicolumn{1}{c}{0.0301} &\multicolumn{1}{c}{0.081} &\multicolumn{1}{c}{0.0485} &\multicolumn{1}{c}{0.0344 (-37\%)} &\multicolumn{1}{c}{0.0040 (-87\%)} &\multicolumn{1}{c}{0.0696 (-14\%)} &\multicolumn{1}{c}{0.0263 (-46\%)} \\
\multirow{1}{*}{Arts} &\multicolumn{1}{c}{0.0946} &\multicolumn{1}{c}{0.0511} &\multicolumn{1}{c}{0.1464} &\multicolumn{1}{c}{0.0811} &\multicolumn{1}{c}{0.0443 (-53\%)} &\multicolumn{1}{c}{0.0067 (-87\%)} &\multicolumn{1}{c}{0.0859 (-41\%)} &\multicolumn{1}{c}{0.0344 (-58\%)} \\
\multirow{1}{*}{Office} &\multicolumn{1}{c}{0.0757} &\multicolumn{1}{c}{0.0424} &\multicolumn{1}{c}{0.1095} &\multicolumn{1}{c}{0.0669} &\multicolumn{1}{c}{0.0391 (-48\%)} &\multicolumn{1}{c}{0.0056 (-87\%)} &\multicolumn{1}{c}{0.0735 (-33\%)} &\multicolumn{1}{c}{0.0309 (-54\%)} \\
\multirow{1}{*}{Games} &\multicolumn{1}{c}{0.05} &\multicolumn{1}{c}{0.0275} &\multicolumn{1}{c}{0.0755} &\multicolumn{1}{c}{0.0444} &\multicolumn{1}{c}{0.0211 (-58\%)} &\multicolumn{1}{c}{0.0015 (-95\%)} &\multicolumn{1}{c}{0.0442 (-41\%)} &\multicolumn{1}{c}{0.0171 (-61\%)} \\
\multirow{1}{*}{Pet} &\multicolumn{1}{c}{0.069} &\multicolumn{1}{c}{0.0318} &\multicolumn{1}{c}{0.1085} &\multicolumn{1}{c}{0.0578} &\multicolumn{1}{c}{0.0391 (-43\%)} &\multicolumn{1}{c}{0.0094 (-70\%)} &\multicolumn{1}{c}{0.0680 (-37\%)} &\multicolumn{1}{c}{0.0317 (-45\%)} \\
\bottomrule
\end{tabular}
}
\caption{Data deduplication. Optimistic evaluations of LIR Baseline (always recommend last-item in the input sequence) on data from~\cite{10.1145/3580305.3599519} (w/o dedup) and our data w/ dedup respectively.}
\label{table:DataDedupe}
\end{center}
\end{table*}

\begin{table*}[!ht]
\begin{center}
\resizebox{0.98\linewidth}{!}{%
\begin{tabular}{cccccccc}
\toprule 
\rowcolor{Gray}
\multirow{1}{*}{} &\multicolumn{1}{c}{}  &\multicolumn{2}{c}{\textsc{Output Matches an Item} (\%)} &\multicolumn{4}{c}{\textsc{Hierarchical Match} - \textsc{Metric}: (\textit{opt})/(\textit{pes}) \& Improv. vs BM25 (\%)} \\
\rowcolor{Gray}
\multirow{1}{*}{Dataset} &\multicolumn{1}{c}{Format Error (\%)} &\multicolumn{1}{c}{All Four Attributes} &\multicolumn{1}{c}{Title Only} &\multicolumn{1}{c}{NDCG@10} &\multicolumn{1}{c}{Recall@1} &\multicolumn{1}{c}{Recall@10} &\multicolumn{1}{c}{MRR} \\
\cmidrule(lr){2-2} \cmidrule(lr){3-4} \cmidrule(lr){5-8}
\multirow{1}{*}{Scientific} &\multicolumn{1}{c}{0.0\%} &\multicolumn{1}{c}{\multirow{1}{*}{82.3\%} } &\multicolumn{1}{c}{\multirow{1}{*}{96.7\%} } &\multicolumn{1}{c}{0.0771 / 0.0725 (-2.1\%)} &\multicolumn{1}{c}{0.0525 / 0.0404 (+3.2\%)} &\multicolumn{1}{c}{0.1054 / 0.1054 (-5.9\%)} &\multicolumn{1}{c}{0.0693 / 0.0632 (-5.2\%)} \\
\multirow{1}{*}{Instruments} &\multicolumn{1}{c}{\multirow{1}{*}{0.001\%} } &\multicolumn{1}{c}{\multirow{1}{*}{97.6\%} } &\multicolumn{1}{c}{\multirow{1}{*}{99.8\%} } &\multicolumn{1}{c}{0.0905 / 0.0901 (-0.4\%)} &\multicolumn{1}{c}{0.0717 / 0.0705 (-0.1\%)} &\multicolumn{1}{c}{0.1151 / 0.1151 (-0.6\%)} &\multicolumn{1}{c}{0.0836 / 0.0830 (-3.2\%)} \\
\multirow{1}{*}{Arts} &\multicolumn{1}{c}{\multirow{1}{*}{0.01\%} } &\multicolumn{1}{c}{\multirow{1}{*}{92.5\%} } &\multicolumn{1}{c}{\multirow{1}{*}{98.9\%} } &\multicolumn{1}{c}{0.0854 / 0.0844 (-1.1\%)} &\multicolumn{1}{c}{0.0636 / 0.0611 (+0.1\%)} &\multicolumn{1}{c}{0.1113 / 0.1113 (-2.3\%)} &\multicolumn{1}{c}{0.0777 / 0.0764 (-4.6\%)} \\
\multirow{1}{*}{Office} &\multicolumn{1}{c}{0.0\%} &\multicolumn{1}{c}{\multirow{1}{*}{84.5\%} } &\multicolumn{1}{c}{\multirow{1}{*}{94.7\%} } &\multicolumn{1}{c}{0.0926 / 0.0916 (-5.2\%)} &\multicolumn{1}{c}{0.0736 / 0.0710 (-3.3\%)} &\multicolumn{1}{c}{0.1127 / 0.1127 (-7.1\%)} &\multicolumn{1}{c}{0.0864 / 0.0851 (-6.6\%) } \\
\multirow{1}{*}{Games} &\multicolumn{1}{c}{\multirow{1}{*}{0.05\%} } &\multicolumn{1}{c}{\multirow{1}{*}{98.0\%} } &\multicolumn{1}{c}{\multirow{1}{*}{99.0\%} } &\multicolumn{1}{c}{0.0582 / 0.0573 (-2.1\%)} &\multicolumn{1}{c}{0.0276 / 0.0257 (-6.8\%)} &\multicolumn{1}{c}{0.0973 / 0.0972 (-1.3\%)} &\multicolumn{1}{c}{0.0467 / 0.0456 (-8.4\%)} \\
\multirow{1}{*}{Pet} &\multicolumn{1}{c}{\multirow{1}{*}{0.03\%} } &\multicolumn{1}{c}{\multirow{1}{*}{97.3\%} } &\multicolumn{1}{c}{\multirow{1}{*}{99.0\%} } &\multicolumn{1}{c}{0.0733 / 0.0690 (-0.4\%)} &\multicolumn{1}{c}{0.0573 / 0.0468 (+0.6\%)} &\multicolumn{1}{c}{0.0924 / 0.0922 (-1.3\%)} &\multicolumn{1}{c}{0.0677 / 0.0620 (-2.8\%)} \\

\bottomrule
\end{tabular}
}
\caption{Hierarchical match results. The relative gains/drops in $\frac{\text{(\textit{opt})}+\text{(\textit{pes})}}{2}$ are also reported against BM25 retrieval.}
\label{table:hierarchical}
\end{center}
\end{table*}

\sparagraph{Further Discussion on User Sequence Dedup} 
Now, we further delve deeper into the issue of duplicate interactions in user sequences.
Although in real world users can make repeated activities, we still posit that the high ratio of consecutive identical interactions in the raw Amazon Review data are not likely to reflect normal/real user behavior (further details in Appendix/Auxiliary Materials). While we \textit{cannot conclude} that data deduplication is `correct' or `wrong', we reckon that the `LIR' baseline results are meaningful and should be reported for model comparison purposes. 
We further run the LIR Baseline on \textbf{1)} data from~\citet{10.1145/3580305.3599519} without dedup and compare with its results on \textbf{2)} our data with dedup. Although the data statistics are slightly different,
the comparisons are still meaningful, since their data sizes are comparable and they are both representative of the full distribution. Table~\ref{table:DataDedupe} shows the optimistic metrics of LIR Baseline that always recommends last item in the input sequence on data w/o dedup (left four columns) and data w/ dedup (right four columns). All metrics suffer sharp declines due to the dedup process. Specifically, Recall@1 drops $60\%$-$95\%$ after removing consecutive identical events. This highlights the importance of user sequence interaction deduplication as a critical step for reliably measuring model quality. 

\rparagraphnodot{Does CALRec Rely on BM25?}
One challenge of using generative LLMs, especially for general-purpose LLMs not tuned with task-specific data, for recommendation is that there is no guarantee that the predicted item exists in the item corpus. Therefore we use BM25, a fuzzy lexical matching algorithm, to rank the items in the item corpus. However, the fine-tuned CALRec models can quickly master the item template format and remember items seen during training, making direct text-entity based retrieval possible without relying on fuzzy text matching methods. This motivates a simple and faster alternative to BM25.

First, we found that nearly all ($>99.9\%$) fine-tuned model outputs adhere to the desired item text format: ``Title: **. Category: **. Brand: **. Price: **.''. We then extract the four attribute entities: title, category, brand, and price, and find that $82\%-98\%$ of model outputs correspond to an item in the item corpus that exactly matches all four attributes, and $95\%-99.8\%$ of title entities extracted from the model outputs exist in the item corpus. These statistics are summarized in Table~\ref{table:hierarchical}. Based on these findings, we propose a heuristic hierarchical matching method. We first search exact matches in the item corpus for all four attributes extracted from model outputs; If no match is found, we search for the exact match of the first three attributes, then the first two, and finally only the title description. Similar to our main experiments, both optimistic and pessimistic metrics are calculated, with results also presented in Table~\ref{table:hierarchical}: we find that our hierarchical matching method has marginally worse performance compared to our BM25-based retrieval, yet it still outperforms other baselines models presented in Table~\ref{table:MainRes}.

\rparagraph{Limitations and Future Work}
Due to the fact that fine-tuned CALRec usually outputs exact item descriptions seen during training, a direct limitation ensues: CALRec barely cannot address the \textit{item cold-start} scenario where the ground-truth item is neither covered in the training set nor in the validation set. We leave addressing this limitation to future work. While previous approaches that also rely on matching LLM-generated item descriptions (e.g., titles) did not provide experimental results on item cold-start cases~\cite{10.1007/978-3-031-56063-7_42,li2023gpt4rec}, we emphasize that our work is the first to identify this issue, which may also be present in these and similar approaches (yet to be verified). We regard this finding a minor contribution of our study.

Second, although CALRec outperforms other baselines in Recall@1, NDCG@10, and MRR@$\infty$ by large margin, its Recall@10 underperforms one of our ID+text baselines. We attribute this to the quasi-greedy nature of temperature sampling (especially when adopting a relatively small temperature value). Unlike embedding-based nearest neighbor search, our approach may not consistently produce the most optimal top-10 item predictions from the entire item corpus. Consequently, although CALRec excels in metrics that emphasize the relevance of the very top item prediction (like Recall@1 and MRR), there is a need for greater diversity in its top-10 item predictions.

Third, in real-world scenarios, when the size of item corpus is extremely large (e.g., billions of items), BM25 retrieval becomes  prohibitively slow and often infeasible. In addition to the hierachical entity matching method discussed earlier in this section, we may leverage CALRec as a reranker within a retrieve-and-rerank framework where any baseline sequential recommendation approach can be used to first derive, for example, top 10K recommendations focused on high recall. Then CALRec's output texts are used to rerank these items. This retrieve-and-rerank paradigm can not only leverage the strengths of CALRec but also circumvents the need for BM25 retrieval on a massive corpus.


\section{Conclusion}
\label{s:conclusion}
We proposed CALRec, a contrastive-learning-assisted two-stage training framework for sequential recommendation based on LLMs, with experiments conducted using the PaLM-2 LLM as the backbone. Our approach also features careful template design inspired from few-shot learning and a novel quasi-round-robin BM25 retrieval approach. Comprehensive experiments on the Amazon Review dataset demonstrate the effectiveness of CALRec, supported by PaLM-2, where it significantly outperforms existing SotA models for sequential recommendation by a considerable margin in most evaluation metrics. We also presented a series of further analyses and revisited the issue of duplicate interactions in user sequences and existing evaluation metrics; we accordingly propose to adopt a training-free text-statistic based `Last Item Repeater' reference baseline for model comparison purposes, and we strongly suggest reporting both optimistic and pessimistic metric scores for all text-based approaches.

\begin{acks}
We thank the following Google colleagues (in alphabetical order) for their technical advice and helpful discussions on this work: Aahil Mehta, Ao Liu, Boaz Cogan, Jianmo Ni, Minh Pham, Nicolas Aagnes, Wang-Cheng Kang, Xin Zhang, Yan Zhu. We also extend our gratitude to anonymous reviewers for their constructive feedback.
\end{acks}

\clearpage
\bibliographystyle{ACM-Reference-Format}
\bibliography{references}

\clearpage
\appendix
\label{s:appendix}
\section{Examples of Data Duplication}
As discussed in \S\ref{s:expsetup}, the raw Amazon Review data contain repeated, non-distinguishable user-item interactions that share identical entity values across all attributes, including item id, user id, timestamp, item metadata, user rating, user review text, etc. An example in the `Scientific' category is as follows:
\begin{lstlisting}[language=json]
{
  "uid": "A3SFSFJZFI0OQN",
  "asin": ["B00002NC3K", "B0001MSC84", "B0001MSC84", "B0001MSC84", "B0001MSC84"],
  "overall": [3, 3, 3, 3, 3],
  "reviewText": [
    "Warped when exposed to the sun.",
    "Warped when exposed to the sun.",
    "Warped when exposed to the sun.",
    "Warped when exposed to the sun.",
    "Warped when exposed to the sun.",
  ],
  "unixReviewTime": [1508544000, 1508544000,  1508544000,  1508544000, 1508544000],
  "title": [
    "Rubbermaid Commercial Products FG263100GRAY Rubbermaid Commercial Round Brute Container Lid, Gray, 32G",
    "Rubbermaid Commercial BRUTE Heavy-Duty Round Waste/Utility Container with Venting Channels, 20-gallon, Gray (FG262000GRAY)",
    "Rubbermaid Commercial BRUTE Heavy-Duty Round Waste/Utility Container with Venting Channels, 20-gallon, Gray (FG262000GRAY)",
    "Rubbermaid Commercial BRUTE Heavy-Duty Round Waste/Utility Container with Venting Channels, 20-gallon, Gray (FG262000GRAY)",
    "Rubbermaid Commercial BRUTE Heavy-Duty Round Waste/Utility Container with Venting Channels, 20-gallon, Gray (FG262000GRAY)",
  ],
}
\end{lstlisting}
In the example provided, all $5$ user-item interactions, except the initial one, are identical. This behavior generally deviates from normal user activities and may potentially confuse or mislead models into recommending repetitive items
. Although duplicate events make up only a small percentage of the dataset (Table~\ref{table:DataStats}), our analysis via the LIR baseline on raw and deduplicated data (Table~\ref{table:DataDedupe}) shows that those duplicates can significantly affect model behavior by offering an overly simplistic pattern to learn from.

\begin{figure}[ht!]
    \centering
    \includegraphics[width=1.0\linewidth]{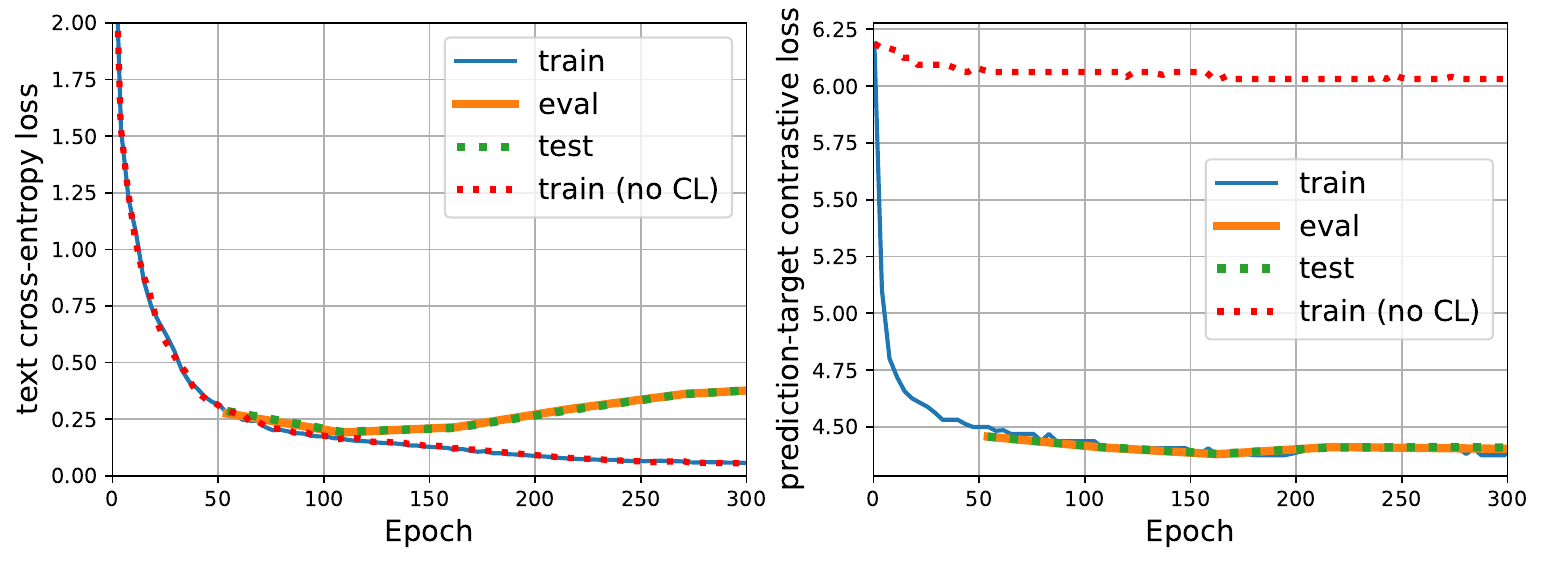}
    \caption{Text generation cross-entropy loss ($\mathcal{L}_{NIG}$, left panel) and prediction-target contrastive loss ($\mathcal{L}_{TT}$, right panel) during Scientific-category-specific fine-tuning without multi-category joint fine-tuning. In both
    panels, the training curves (red-dotted) without contrastive alignment are also plotted for
    comparison.}
    \label{fig:loss}
\end{figure}

\section{Additional Discussion on Contrastive Alignment}
\label{appendix:contrastive}
Fig.~\ref{fig:loss} shows the loss profiles during category-specific fine-tuning on the Scientific category without Stage \RN{1} multi-category joint fine-tuning. The left panel shows that overfitting happens in text generation task after circa $100$ epochs. The right panel shows that train, evaluation and test contrastive loss between prediction and target drops quickly from $6.2$ to $4.4$ and remains relatively stable. At batch size $512$, the initial $\mathcal{L}_{TT}\approx6.2$ means the average ratio between $\exp(\cos(\mathbf{v}_i^{T|U}, \mathbf{v}_i^T)/\tau_c)$ and $\exp(\cos(\mathbf{v}_j^{T|U}, \mathbf{v}_i^T)/\tau_c), i\neq j$ is about $512\times\exp(-6.2)=1.04$ at the beginning of the training process. Thus $\cos(\mathbf{v}_i^{T|U}, \mathbf{v}_i^T)-\cos(\mathbf{v}_j^{T|U}, \mathbf{v}_i^T)=\ln(1.04)\tau_c\approx0.02\ll1$, suggesting that without fine-tuning, the raw pretrained LLMs, although able to make some prediction (Table~\ref{table:rawulm}), does not yield embeddings for predicted items (in the user tower) that are more aligned with embeddings of the ground truth items (in the item tower) than the embeddings from random users do. After fine-tuning, $\mathcal{L}_{TT}\approx4.4$ and the same ratio increases to $512\times\exp(-4.4)=6.3$, corresponding to $\cos(\mathbf{v}_i^{T|U}, \mathbf{v}_i^T) - \cos(\mathbf{v}_j^{T|U}, \mathbf{v}_i^T)\approx0.92$ which is close to $1$, indicating that the embeddings
 are aligned very well geometrically.
On the contrary, without contrastive alignment, the prediction-target contrastive loss $\mathcal{L}_{TT}$ only drops roughly from $6.4$ to $6.0$ as the model learns to generate the correct next-item prediction as a pure text generation task. Fig.~\ref{fig:loss} demonstrates that the contrastive loss can efficiently align the embeddings from our two towers.

\end{document}